\pdfoutput=1
\documentclass[11pt,twoside,a4paper]{article}

\usepackage[utf8]{inputenc}
\usepackage[hang,normal,footnotesize]{caption}
\usepackage{subcaption}
\usepackage{wrapfig}
\usepackage[sort&compress,numbers,colon,merge]{natbib}
\usepackage{a4wide}
\usepackage{fancyhdr}
\usepackage{graphicx}
\usepackage{url}
\usepackage{amsfonts}
\usepackage{amsmath,amssymb,amsthm}

\usepackage{mathrsfs}

\usepackage{multirow}
\usepackage{color}
\definecolor{orange}{rgb}{1,0.5,0}
\usepackage[colorlinks]{hyperref}
\usepackage{cancel}
\usepackage{bbm}
\usepackage{ulem}

\usepackage{booktabs}
\hypersetup{
pdfauthor={Martin Holthausen, Manfred Lindner and Michael A. Schmidt},
pdftitle={Lepton flavor at the electroweak scale: A complete A4 model}
}

\makeatletter
\renewcommand{\fnum@table}{\textbf{\tablename~\thetable}}
\renewcommand{\fnum@figure}{\textbf{\figurename~\thefigure}}
\makeatother

\numberwithin{equation}{section}

\newcommand{\re}{\ensuremath{\mathrm{Re}}}
\newcommand{\im}{\ensuremath{\mathrm{Im}}}

\newcommand{\diag}{\ensuremath{\mathrm{diag}}}
\newcommand{\I}{\ensuremath{\mathrm{i}}}

\newcommand{\keV}{\ensuremath{\,\mathrm{keV}}}
\newcommand{\MeV}{\ensuremath{\,\mathrm{MeV}}}
\newcommand{\GeV}{\ensuremath{\,\mathrm{GeV}}}
\newcommand{\TeV}{\ensuremath{\,\mathrm{TeV}}}

\newcommand{\braket}[1]{\ensuremath{\left<#1\right>}}
\newcommand{\ev}[1]{\ensuremath{\left\langle#1\right\rangle}}
\newcommand{\VEV}[1]{\ensuremath{\left\langle#1\right\rangle}}
\newcommand{\vev}[1]{\ensuremath{\left\langle#1\right\rangle}}
\newcommand{\hc}{\ensuremath{\text{h.c.}}}
\newcommand{\Order}[1]{\ensuremath{\mathcal{O}(#1)}}
\newcommand{\be}{\begin{equation}}
\newcommand{\ee}{\end{equation}}
\newcommand{\ba}{\begin{eqnarray}}
\newcommand{\ea}{\end{eqnarray}}
\newcommand{\SU}[1]{\ensuremath{\mathrm{SU}(#1)}}
\newcommand{\U}[1]{\ensuremath{\mathrm{U}(#1)}}

\newcommand{\Eqref}[1]{Eq.~(\ref{#1})}

\newcommand{\Figref}[1]{Fig.~\ref{#1}}
\newcommand{\Tabref}[1]{Tab.~\ref{#1}}
\newcommand{\Secref}[1]{sec.~\ref{#1}}
\newcommand{\Appref}[1]{appendix \ref{#1}}

\renewcommand{\subsubsection}[1]{\vspace{1ex}\mathversion{bold}{\bf #1:}\mathversion{normal}}

\newcommand{\Rep}[1]{\ensuremath{\underline{\mbox{\textbf{#1}}}}}
\newcommand{\MoreRep}[2]{\ensuremath{\underline{\mbox{\textbf{#1}}} _{\mbox{\textbf{#2}}}}}
\newcommand{\SG}[2]{\ensuremath{\mathrm{SG}(#1,#2)}}

\newcommand{\abs}[1]{\ensuremath{\left\vert#1\right\vert}}

\newcommand{\ra}[1]{\renewcommand{\arraystretch}{#1}}

\begin{document}
\allowdisplaybreaks[1]

%%%%%%%%%%%%%%%%%%
%%  TITLE PAGE  %%
%%%%%%%%%%%%%%%%%%

\begin{titlepage}

\begin{center}
{\huge\sffamily\bfseries\mathversion{bold} 
Lepton flavor at the electroweak scale:\\ A complete $A_4$ model
\mathversion{normal}
}
\\[10mm]
{\large
Martin Holthausen\footnote{\texttt{martin.holthausen@mpi-hd.mpg.de}}$^{(a)}$, Manfred Lindner\footnote{\texttt{lindner@mpi-hd.mpg.de}}$^{(a)}$ and 
Michael A.~Schmidt\footnote{\texttt{michael.schmidt@unimelb.edu.au}}$^{(b)}$}
\\[5mm]
{\small\textit{$^{(a)}$
Max-Planck Institut f\"ur Kernphysik, Saupfercheckweg 1, 69117
Heidelberg, Germany
}}\\
{\small\textit{$^{(b)}$
ARC Centre of Excellence for Particle Physics at the Terascale,
School of Physics, The University of Melbourne, Victoria 3010, Australia
}}

\end{center}
\vspace*{1.0cm}
\date{\today}

\begin{abstract}
\noindent
Apparent regularities in fermion masses and mixings are often associated with physics at a high flavor scale, especially in the context of discrete flavor symmetries. One of the main reasons for that is that the correct vacuum alignment requires usually some high scale mechanism to be phenomenologically acceptable. Contrary to this expectation, we present in this paper a renormalizable radiative neutrino mass model with an $A_4$ flavor symmetry in the lepton sector, which is broken at the electroweak scale. For that we use a novel way to achieve the VEV alignment via an extended symmetry in the flavon potential proposed before by two of the authors. We discuss various phenomenological consequences for the lepton sector and show how the remnants of the flavor symmetry suppress large lepton flavor violating processes. The model naturally includes a dark matter candidate, whose phenomenology we outline. Finally, we sketch possible extensions to the quark sector and discuss its implications for the LHC, especially how an enhanced diphoton rate for the resonance at 125 GeV can be explained within this model.
\end{abstract}

\end{titlepage}

\setcounter{footnote}{0}

\section{Introduction}
Much progress has been achieved in the field of particle physics during the last year. First, the last missing mixing angle $\theta_{13}$ of the Standard Model(SM) with massive neutrinos has been measured\cite{An:2012eh,*Ahn:2012nd} to be $8^\circ$ after first hints in 2011~\cite{Abe:2011sj,*Adamson:2011qu,*Abe:2011fz} and recently an excess consistent with a SM Higgs has been observed at $126.0 \pm 0.4(\mathrm{stat.}) \pm 0.4(\mathrm{sys.})$ GeV by ATLAS~\cite{:2012gk,ATLAS-CONF-2012-162} and at $125.8\pm 0.4(\mathrm{stat.}) \pm 0.4(\mathrm{sys.})$ GeV by CMS~\cite{:2012gu,CMS-HIG-12-045}. 

Let us first discuss the implications of the large mixing angle $\theta_{13}$. Much of the work in the neutrino sector has been aimed at explaining tiny values of $\theta_{13}$ as deviations from a tri-bi-maximal(TBM) mixing structure using flavor symmetries, but this scenario is now implausible due to the sizable value of $\theta_{13}$\footnote{See also the recent global fits to neutrino oscillations~\cite{Tortola:2012te,Fogli:2012ua,*GonzalezGarcia:2012sz}. }.
 So maybe there exists no symmetry that is connected to the regularities in the fermion parameters. It could rather be that mixing angles are determined at a high scale from some (quasi-)random mechanism. Indeed if one randomly draws unitary $3\times 3$ matrices with a  probability measure given by the Haar measure of \U{3}, i.e. the unique measure that is invariant under a change of basis for the three generations, one finds a probability of $44 \%$ for nature to have taken a more "unusual" choice~\cite{de-Gouvea:2012fj}. This cannot be interpreted, however, as an indication in favor of anarchy~\cite{de-Gouvea:2003uq,Hall:2000qy}, as the sample (3 mixing angles and one mass ratio) is clearly too small to reconstruct the probability measure to any degree of certainty~\cite{Espinosa:2003yq}. The only statement one can make is that the (very limited) data cannot rule out the anarchy hypothesis. For any values of the mixing angles one can always find a flavor model that is in better agreement with the data\footnote{This can be done without increasing the degree of complexity of the model\cite{Altarelli:2012lr}.}.

Another option is that flavor symmetries are realized in a different way. One route is to think of solutions that do not predict TBM. Such models are usually implemented at high scales and give precise predictions for the leptonic mixing angles in the experimentally allowed regions. These models might then be falsified in the same way that the models that give tri-bi-maximal mixing have been ruled out, i.e. by a further refinement of the experimental determination of these angles. This seems to be the only fruitful direction for models that explain flavor at high energy scales such as the see-saw or GUT scale, because mixing angles are generically the only experimentally testable predictions of such models. An example are models based on $\Delta(96)$\cite{de-Adelhart-Toorop:2012fv,Ding:2012fr,King:2012fk}. 

An interesting question is if flavor symmetries could even be realized at low scales\cite{Ma:2001fk,Ma:2006km,Kubo:2006kx,Hirsch:2009mx,Ibanez:2009du,Ma:2010gs,Adelhart-Toorop:2010fk,Adelhart-Toorop:2010uq,Bhattacharyya:2010hp,Cao:2011df,Cao:2010mp,Bhattacharyya:2012ze,*Bhattacharyya:2012pi}. If this is viable then such models can be tested by additional observables. Such observables typically include rare lepton flavor violating(LFV) decays of leptons and mesons and --ideally-- a direct experimental access to the very fields that mediate the flavor symmetry breaking. Typically such models will feature extended Higgs sectors but will not uniquely determine the mixing angles; they will, however, rather give relations among the deviations from patterns such as tri-bi-maximal mixing. 

In this work we implement a model based on a flavor symmetry at the electroweak scale and show that it leads to additional phenomenological effects that will become testable in the future.

Many authors have noted that quite generic corrections in the neutrino sector within models based on the symmetry group $A_4$ \cite{Ma:2001fk,Babu:2002dz,*Ma:2004zv,*Altarelli:2005yp,*Babu:2005se,*Altarelli:2005yx,*He:2006dk} may lead to corrections to the leptonic mixing angles that are in agreement with the experimental data~\cite{Brahmachari:2008fn,*Ma:2011yi,*Shimizu:2011xg,*King:2011zj,*Antusch:2011ic,*King:2011kk,*Ahn:2012cg,*Ishimori:2012fg,*Ma:2012ez,*Branco:2012vs,*Chen:2012st}. It should be noted, however, that all such models need to break the symmetry group in a particular way to different subgroups in the charged lepton and neutrino sectors, a vacuum configuration that cannot be obtained from a straightforward minimization of the potential but something that rather needs a special dynamical mechanism to achieve it. 
The two most commonly used mechanisms are either based on (i) continuous R-symmetries in supersymmetry or (ii) extra-dimensions. The supersymmetric models only work in the limit where the supersymmetry breaking scale is below the scale of flavor symmetry breaking and the scale of flavor symmetry breaking is thus unobservabely high. The only experimentally verifiable prediction of such models seem to be correlations in the deviations from TBM, e.g. the trimaximal mixing pattern that predicts $
a\approx-\frac12 r\cos\delta $ and  $ s\approx0
\label{eq:tri-max-sumrule}
$
where~\cite{King:2007pr,*Grimus:2008tt,*Albright:2008rp,*Ge:2011ih,*Ge:2011qn,*Hernandez:2012ra,*Hernandez:2012sk}
\begin{align}\label{eq:devTBM}
\sin\theta_{13}&=\frac{r}{\sqrt{2}},&
\sin\theta_{12}&=\frac{1}{\sqrt{3}}\left(1+s\right),&
\sin\theta_{23}&=\frac{1}{\sqrt{2}}\left(1+a\right)\;.
\end{align}

As these models tend to be quite baroque, since they involve a large number of driving fields etc., this one prediction seems to be a rather poor showing compared to the model building effort involved. The second possibility using extra-dimensions is possible, but we here want to focus on non-supersymmetric models in four dimension, a possibility that seems to be favored by the experimental data.  

In an earlier work~\cite{Holthausen:2011vd}, two of us(MH and MS) have studied a possibility to obtain the vacuum alignment that seems to be quite close to the general spirit of discrete flavor model building: to get a predictive model one needs to have unbroken discrete remnant symmetries in the neutrino and charged lepton sectors that do not commute with each other. The way this is usually  realized is that symmetries, particle content and vacuum structure are selected such that these remnant symmetries are  realized as accidental symmetries of the leading-order (LO) mass matrices that are broken at next-to-leading order. The idea of~\cite{Holthausen:2011vd}, following earlier work by Babu and Gabriel~\cite{Babu:2010bx}, was to have an extended flavor group $G$ and a particle content such that an accidental symmetry $G\times A_4$ arises at the renormalizable level in the scalar potential that allows for the desired vacuum configuration. More precisely, the scalars $\chi_i$ breaking the extended flavor group $G$ in the charged lepton sector transform under the $A_4$ only, while the scalars $\phi_i$ breaking the extended flavor group $G$ transform under the full group $G$. The symmetry is chosen such that the scalar potential does not contain operators with non-trivial contractions of $\chi_i$ with $\phi_i$, i.e. there are only contractions of the form $(\chi_i\chi_j)_{\MoreRep{1}{1}}(\phi_i\phi_j)_{\MoreRep{1}{1}}$.
The smallest symmetry group that  realizes such a structure for $A_4$ is $G= Q_8\rtimes A_4$ and since the symmetry breaking does not need any special additional ingredient, there is no immediate theoretical obstacle to have the flavor symmetry breaking scale at the electroweak scale. 

In this work we therefore implement the model in~\cite{Holthausen:2011vd} at the electroweak scale. The outline of the paper is as follows: We introduce the model and the exact symmetry breaking pattern in \Secref{sec:model}. Without the introduction of any additional symmetries, we show in \Secref{sec:neutrino} that neutrino masses are generated at one-loop level and that the neutrino mass matrix is determined by five physical parameters. We discuss the predictions for neutrino oscillation observables that follow from this structure. In \Secref{sec:LFVprocesses}, we discuss constraints from lepton-flavor violating rare decays. In \Secref{sec:DM}, we show that the model contains a dark matter candidate and discuss its phenomenology. Three simple possible extensions to the quark sector are discussed in \Secref{sec:quarks}. Finally, in \Secref{sec:collider} we discuss the implications of direct collider searches and the the recent observation of a Higgs-like boson at 125 GeV and then we conclude in \Secref{sec:conclusions}.

%%%%%%%%%%%%%%%%%%%%%%%%%%%%%%%%%%%%%%%%%%%%%%%%%%%%%%%%5
%%%%%%%%%%%%%%%%%%%%%%%%%%%%%%%%%%%%%%%%%%%%%%%%%%%%%%%%%%%%%%%%%%%55
%%%%%%%%%%%%%%%%%%%%%%%%%%%%%%%%%%%%%%%%%%%%%%%%%%%%%%%%%%%%%%%%%%%55
%\input{model-v1}
\section{Model and Symmetry Breaking}
\label{sec:model}
We utilize the symmetry $Q_8\rtimes A_4$ introduced in \cite{Holthausen:2011vd}, which allows for natural vacuum alignment, and implement a model describing the lepton sector at the electroweak scale. Hence, we promote the flavon fields of~\cite{Holthausen:2011vd} that couple to the charged lepton sector to EW Higgs doublets.
The particle content of the lepton sector is given in \Tabref{tab:partcontent-EWscale}. The vacuum configuration 
\begin{align}
\ev{\chi_i}=\left( \begin{array}{c}0\\ \frac{v}{\sqrt{6}} \end{array}\right), \qquad \langle\phi_1\rangle =\frac{1}{\sqrt{2}}(a,a,b,-b)^T,\qquad \langle\phi_2\rangle &=\frac{1}{\sqrt{2}}(c,c,d,-d)^T
\label{eq:vac-conf}
\end{align}
 can be naturally obtained from the most general scalar potential following the discussion in \cite{Holthausen:2011vd}. As the discussion is very similar to the one given there, we relegate it to \Appref{sec:vacuum-alignment-app}, where also the scalar mass spectrum is discussed. However, let us briefly recall the salient features of the vacuum expectation value (VEV) configuration (\ref{eq:vac-conf}): the scalar singlets $\phi_1$ and $\phi_2$ break the symmetry group to the subgroup $\ev{S\vert S^2=E}\cong Z_2$ and the EW doublets $\chi$ break the discrete symmetry group down to the subgroup $\ev{T\vert T^3=E}\cong Z_3$, while simultaneously breaking the electroweak gauge group $\SU{2}_L\times \U{1}_Y$ down to the electromagnetic $\U{1}_{\mathrm{em}}$. The  normalization is chosen such that $\sum_{i}v_i^2= v^2=(\sqrt{2}G_F)^{-1}= (246 \GeV)^2$, in accordance with our earlier definition.
Because of the unbroken $Z_3$ symmetry in the charged lepton sector, it is useful to go to a basis~\cite{Ma:2001fk,Ma:2010gs,Adelhart-Toorop:2010fk,Adelhart-Toorop:2010uq}
\begin{align}
\left(H, {\varphi^{\prime}}, {\varphi^{\prime \prime}} \right)^T=\Omega_T^\dagger \chi\sim (1,\omega^2,\omega), \qquad \left(L_e, L_\mu, L_\tau \right)^T=\Omega_T^\dagger L\sim (1,\omega^2,\omega),
\label{eq:physical-basis-triality}
\end{align}
where this symmetry is represented diagonally and $\Omega_T$ is defined by
\begin{align}\label{eq:SigmaTdef}
\Omega_T\equiv\frac{1}{\sqrt{3}}
\left(
\begin{array}{ccc}
 1 & 1 & 1 \\
 1 & \omega ^2  & \omega  \\
 1 & \omega  & \omega ^2 
\end{array}
\right)\;.
\end{align}
We have indicated the transformation properties under the unbroken subgroup $\vev{T}\cong Z_3$ under which $(e^c,\mu^c,\tau^c)$ transform as $(1,\omega,\omega^2)$. This has been denoted flavor triality in~\cite{Ma:2010gs}. In this basis the vacuum configuration (\ref{eq:vac-conf}) implies that only the field $H$ acquires a VEV $\vev{H}=\left(0,v/\sqrt{2} \right)^T,$ 
while $\varphi^\prime$ and $\varphi^{\prime \prime}$ are inert doublets (and thus do not obtain a VEV).
The potential for the electroweak doublets $\chi$ is given by
\begin{align}
V_\chi(\chi)&=\mu^2_3 \chi^\dagger\chi+ \hspace{-.5cm}\sum_{r=\MoreRep{1}{1,2},\MoreRep{3}{1S,1A}}\hspace{-.5cm}\lambda_{\chi r} (\chi^\dagger \chi)_{r}(\chi^\dagger \chi)_{r^*}+\lambda_{\chi A}\im \left[ (\chi^\dagger \chi)_{\MoreRep{3}{1S}}(\chi^\dagger \chi)_{\MoreRep{3}{1A}} \right],
\label{eq:chi-potential}
\end{align}
and after symmetry breaking the nine physical scalars contained in $\chi$ arrange themselves in the following multiplets under the remnant $\U{1}\times Z_3$ symmetry. There is one real scalar $h=\sqrt{2} \re H^0 $ with mass
\begin{align}
m^2_h=\frac{2}{9} \left(3 \lambda _{\text{$\chi $ }1_1}+\sqrt{3} \lambda _{\text{$\chi $ }3_{1,S}}\right)v^2
\end{align}
that plays the role of the Standard Model Higgs. Note that since this scalar is a complete singlet under all remnant symmetries, it can in principle mix with components of $\phi_1$ and $\phi_2$ that transform in the same way. This is discussed in detail in \Eqref{eq:Higss-mixing-angles} in the appendix and in the following we will for the most part assume the mixing to be small enough to treat $h$ as a mass eigenstate.
\begin{table}
\centering
\ra{1.2}
\begin{tabular}{lcccccccccccc}\toprule
 & $L$ &$e^c$ &$\mu^c$ &$\tau^c$ &$\chi$&$\phi_1$&$\phi_2$&&$S$ & $\eta_1$&$\eta_2$&$\eta_3$ \\ \cmidrule{1-8}\cmidrule{10-13}
$Q_8\rtimes A_4$& $\MoreRep{3}{1}$&$\MoreRep{1}{1}$&$\MoreRep{1}{2}$&$\MoreRep{1}{3}$&$\MoreRep{3}{1}$&$\MoreRep{4}{1}$&$\MoreRep{4}{1}$&&$\MoreRep{3}{2}$&$\MoreRep{3}{5}$&$\MoreRep{3}{4}$&$\MoreRep{3}{5}$\\
$Z_4$&$\I$&$-\I$ &$-\I$&$-\I$&$1$&$1$&$-1$&&$-1$&$\I$&$\I$&$-\I$\\
$\SU{2}_L$&$2$&$1$ &$1$&$1$&$2$&$1$&$1$&&$1$&$2$&$2$&$2$\\
$\U{1}_Y$&$-1/2$&$1$ &$1$&$1$&$1/2$&$0$&$0$&&$0$&$1/2$&$1/2$&$1/2$\\
\bottomrule
\end{tabular}
\caption{Particle content of the minimal model that  realizes flavor symmetry breaking at the electroweak scale. The flavon $\chi$ contains the Higgs field and ties the electroweak to the flavor breaking scale. The scalars $\eta_i$ and fermionic multiplet $S$ are needed for one-loop generation of neutrino masses.\label{tab:partcontent-EWscale}}
\end{table}
\ra{1}

The next four degrees of freedom are  in the charged scalars $\varphi^{\prime+}$ and $\varphi^{\prime \prime +}$ that transform as $(1,\omega^2)$ and $(1,\omega)$ under $\U{1}\times Z_3$, respectively, and have the masses
 \begin{align}
m^2_{\varphi^{\prime+}} 
=\frac{v^2}{12}\left(
  -2 \sqrt{3} \lambda _{\text{$\chi $ }3_{1,S}}-\lambda_{\chi A}
 \right) ,\qquad    m_{\varphi^{\prime \prime+}}
 =\frac{v^2}{12}\left(  -2 \sqrt{3} \lambda _{\text{$\chi $ }3_{1,S}}+\lambda_{\chi A}
 \right) .
 \label{eq:masses-charged}
 \end{align}
The final four real scalars sit in the two complex neutral scalars $\varphi^{\prime 0}$ and ${\varphi^{\prime\prime 0}}^*$, that both transform as $(0,\omega^2)$ and the mass eigenstates are given by the neutral scalars 
\begin{equation}\label{eq:mixVarphiNeutral}
\left(\begin{array}{c}{\Phi_1}\\ {\Phi_2} \end{array} \right)= U_\varphi \left(\begin{array}{c}\varphi^{\prime 0}\\{\varphi^{\prime\prime 0}}^* \end{array} \right)\equiv\left(\begin{array}{cc} \cos \alpha &\sin \alpha\\ -\sin \alpha & \cos \alpha \end{array}\right) \left(\begin{array}{c}\varphi^{\prime 0}\\{\varphi^{\prime\prime 0}}^* \end{array} \right)
\end{equation}
the mixing angle $\alpha$ and their masses may be written as 
\begin{align}
\tan 2 \alpha &=\frac{6 \lambda _{\text{$\chi $ }1_2}+\sqrt{3} \left(3 \lambda _{\text{$\chi
   $ }3_{1,A}}+\lambda _{\text{$\chi $ }3_{1,S}}\right)}{6 \text{$\lambda_{\chi A};
 $}}\;,\\
m^2_{{\Phi_1}}+m_{{\Phi_2}}^2&=-2 \tan (2 \alpha )
   \left(m^2_{\varphi^{\prime \prime+}} -m^2_{\varphi^{\prime+}} \right)+m^2_{\varphi^{\prime \prime+}} +m^2_{\varphi^{\prime+}} -
   \frac{v^2 \lambda _{\text{$\chi $ }3_{1,A}}}{\sqrt{3}}\;,\\   
m^2_{{\Phi_1}}-m_{{\Phi_2}}^2&=2 \abs{\sec (2 \alpha ) }\abs{m^2_{\varphi^{\prime \prime+}} -m^2_{\varphi^{\prime+}} }.
\end{align}
The mass spectra for the other scalars can be found in \Appref{sec:scalar-spectrum}. Two comments are in order here: (i) in the potential \eqref{eq:chi-potential} there is only one mass term for the three doublets. Using the minimization conditions, the mass term can be swapped for the Higgs VEV $v$ and therefore (ii) all of the squared scalar masses are given as a product of dimensionless scalar couplings times $v^2$. The additional scalar masses may therefore not be arbitrarily large. Note that in usual multi-Higgs doublet models each doublet has its own mass term and therefore there is always a decoupling limit where all non-SM particles are unobservabely heavy. Such a setup is therefore directly testable at colliders, as we will study in \Secref{sec:collider}. However, before discussing this, we show that the model accomplishes (i) the description of the (lepton) flavor structure in terms of a small number of parameters and (ii) the protection against bounds on new physics from flavor observables such as lepton flavor violating processes.

%%%%%%%%%%%%%%%%%%%%%%%%%%%%%%%%%%%%%%%%%%%%%%%%%%%%%%%%%%%%%%%%%%%55
%%%%%%%%%%%%%%%%%%%%%%%%%%%%%%%%%%%%%%%%%%%%%%%%%%%%%%%%%%%%%%%%%%%55
%%%%%%%%%%%%%%%%%%%%%%%%%%%%%%%%%%%%%%%%%%%%%%%%%%%%%%%%%%%%%%%%%%%55

%\input{lepton-flavour-structure-v1}
\section{Lepton Flavor Structure}
\label{sec:neutrino}
In this section we discuss the one-loop generation of neutrino masses and phenomenological implications of the predicted flavor structure.
\addtocontents{toc}{\protect\setcounter{tocdepth}{1}}
\subsection{Lepton Masses}
\begin{figure}[tb]
\centering
\includegraphics[width=.6\textwidth]{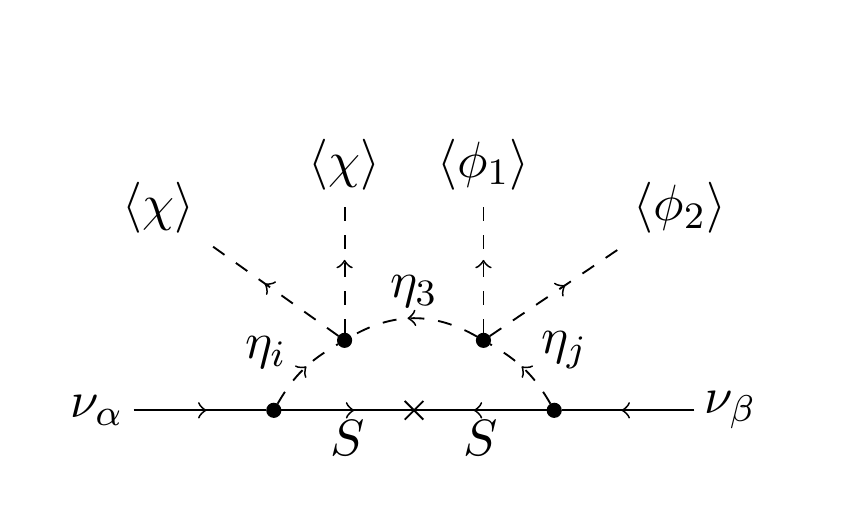}
\caption{Neutrino mass generation at one loop. \label{fig:OneLoopNeutrinoMass}}
\end{figure}

\addtocontents{toc}{\protect\setcounter{tocdepth}{2}}
The charged lepton sector is described by
\begin{equation}
-\mathcal{L}_e = y_e L \tilde{\chi} e^c 
+y_\mu L \tilde{\chi} \mu^c  
+y_\tau L \tilde{\chi} \tau^c   +\hc\; ,
\label{eq:chargedlepton-EW}
\end{equation}
where $\tilde{\chi}=\I \sigma_2 \chi^*$ and here and in the following we do not specifically indicate the contractions if there is only one invariant that can be formed out of the particle content of the operator. In the physical basis of \Eqref{eq:physical-basis-triality} this term reads
\begin{align}
-\mathcal{L}_e =& \tilde{H}\left(y_e L_e e^c +y_\mu L_\mu \mu^c+y_\tau L_\tau \tau^c\right)\
+\tilde{\varphi}'\left(y_e L_\mu e^c +y_\mu L_\tau \mu^c+y_\tau L_e \tau^c\right)\nonumber\\
&+\tilde{\varphi}''\left(y_e L_\tau e^c +y_\mu L_e \mu^c+y_\tau L_\mu \tau^c\right)\; +\hc
\end{align}
and we thus see that $H$ couples diagonally to leptons while $\varphi^{\prime}$ and $\varphi^{\prime \prime}$ do not. Note that here the mass terms are generated by dimension four Yukawa couplings and there is therefore no need for a complicated UV completion. The mass matrix is thus given by
\begin{align}
M_E&=\frac{v}{\sqrt{2}} \Omega_T^*\diag(y_e,y_\mu,y_\tau)\, ,
\end{align}
with $\Omega_T$ given in \Eqref{eq:SigmaTdef}. Neutrino masses are generated at one loop level, through the interactions with the fermionic singlets $S$ and the scalar doublets $\eta$, as shown in \Figref{fig:OneLoopNeutrinoMass}. The couplings of $S$ are given by
\begin{align}\label{eq:YukNu}
\mathcal{L}_\nu= h_1 L \eta_1 S+h_2 L \eta_2 S  +\sqrt{3}\,M_S S S +\hc\;.
\end{align}
The factor of $\sqrt{3}$ cancels a factor coming from the  normalization of Clebsch-Gordon coefficients. In order to calculate the neutrino mass matrix, we have to determine the mass matrix of the neutral components of $\eta_1$, $\eta_2$ and $\eta_3$. To shorten the notation we define the doublet $\hat \eta_J$ to be the $J$-th component of the 9 component vector $\hat \eta=(\eta_1,\eta_2,\eta_3)$ and real scalar field $\hat \eta_k^0$ to be the $k$-th component of $(\sqrt{2} \re \hat \eta^0, \sqrt{2} \im \hat \eta^0 )$.
Besides the direct mass terms
\begin{equation}\label{eq:Veta2}
\left(M_{\eta^0}^2\right)_{ij}=\frac{\partial^2 V_{\eta_i}^{(2)}}{\partial \hat\eta_i^0\partial \hat\eta_j^0} \qquad \mathrm{with} \qquad V_{\eta_i}^{(2)}=\sum_{i=1,2,3}  \sqrt{3}\,M_i^2   \eta_i^\dagger \eta_i\;, 
\end{equation}
there are couplings which give off-diagonal contributions 
\begin{align}
\left(\delta M_{\eta^0}^2\right)_{ij}=\vev{\frac{\partial^2 \delta V_{\eta_i}^{(2)}}{\partial \hat\eta_i^0\partial \hat\eta_j^0}}
\end{align}
to the mass matrix. Such interactions are needed to generate neutrino masses and the relevant ones can be determined from symmetry considerations\footnote{The complete expression for $\delta V_{\eta_i}^{(2)}$ can be found in the appendix in Eq.~\eqref{eq:deltaVeta}. Here, we only present the parts which are relevant for neutrino masses.}. Any contribution to neutrino mass has to be proportional to
\begin{itemize}
\item $M_S$, which breaks the generalized lepton number $L\rightarrow e^{\I \alpha} L, S\rightarrow e^{-\I \alpha } S$
\item either of the couplings $\lambda_{1}$ or $\lambda_{2}$, defined by\footnote{We can set a number of complex parameters real by phase redefinitions. We set $y_e$, $y_\mu$, $y_\tau$, $h_1$, $h_2$, $M_S$, $\lambda_1$, $\lambda_3$, $\lambda_4$ real by rotating $\ell^c, L, \eta_2, S,\chi, \eta_1, \eta_3$, respectively, and display the phase of $\lambda_2$ explicitly.}
\begin{align}\label{eq:Vetachi}
V_\mathrm{\eta,\chi}=&
\lambda_{1} (\chi^T \sigma_2 \vec{\sigma}  \chi)_{\MoreRep{1}{1}} (\eta_1^T \sigma_2 \vec{\sigma}  \eta_3)^*_{\MoreRep{1}{1}}
+\lambda_{2} e^{\I \alpha_{\lambda}}(\chi^T \sigma_2 \vec{\sigma}  \chi)_{\MoreRep{3}{1}} (\eta_2^T \sigma_2 \vec{\sigma}  \eta_3)^*_{\MoreRep{3}{1}}+\hc\;,
\end{align}
which break the generalized lepton number $L\rightarrow e^{\I \alpha} L, \eta_i\rightarrow e^{-\I \alpha } \eta_i$, 
\item and $\lambda_{3}$ or $\lambda_4$ defined by\footnote{The contractions $(\chi^T \sigma_2 \vec{\sigma}  \chi)_{\MoreRep{1}{2,3}} $ vanish in the vacuum given in \Eqref{eq:vac-conf} and thus do not contribute to the masses here, because the $Z_3$ symmetry  generated by $T$ is conserved by $\vev{\chi}$.}
\begin{align}\label{eq:Vetaphi}
V_\mathrm{\eta,\phi}=&
\lambda_3 (\phi_1 \phi_2)_{\MoreRep{1}{1}} (\eta_3^\dagger \eta_1)_{\MoreRep{1}{1}}
+\lambda_4(\phi_1 \phi_2)_{\MoreRep{3}{1}} (\eta_3^\dagger \eta_2)_{\MoreRep{3}{1}}+\hc\;,
\end{align}
which couples to the $Z_4$-breaking VEV of $\phi_2$. 
\end{itemize}
The built-in multiple protection of the neutrino mass operator thus necessitates the large number of couplings involved in neutrino mass generation, and thus a large potential for suppression beyond the naive factor of $1/(16 \pi^2)$ from the loop integral.
For simplicity, we assume that the direct mass terms $M_i$ dominate over all other contributions; this is in fact a necessary condition to have a predictive theory of flavor. Hence, we can approximate the propagator as
\begin{align}\label{eq:approx}
\left[k^2-(M_{\eta^0}^2+\delta M^2_{\eta^0}) \right]^{-1}=(k^2-M_{\eta^0}^2)^{-1}+(k^2-M_{\eta^0}^2)^{-1}\delta M^2_{\eta^0}(k^2-M_{\eta^0}^2)^{-1},
\end{align}
where $M_{\eta^0}^2$ is diagonal, and treat the mixing between the different components of $\eta_i$ by mass insertions $\delta M^2_{\eta^0}$. The evaluation of the one loop diagram leads to
\begin{multline}
\left({M_\nu}\right)_{\alpha \beta}=\sum_{i=1}^{3}\sum_{I,J,M=1}^{18} h_{\alpha i I} h_{\beta i J} M_S \left(\delta M^2_{\eta^0}\right)_{IM}  \left(\delta M^2_{\eta^0}\right)_{MJ}\\
 I\left(\left(M^2_{\eta^0}\right)_{II}^{\frac12},\left(M^2_{\eta^0}\right)_{JJ}^{\frac12},\left(M^2_{\eta^0}\right)_{MM}^{\frac12},M_S \right)
\end{multline}
where the Yukawa couplings $h_{ikJ}$ depend on the two couplings $h_{1,2}$ given in Eq.~(\ref{eq:YukNu}) via 
\begin{equation}
h_{\alpha kJ}=\frac{\partial \mathcal{L}_\nu}{\partial L_\alpha \partial S_k\partial {\hat\eta}_J}
\end{equation}
and the loop integral is given by\footnote{Note that the renormalization scale $\mu$ drops out of the sum; it is displayed here to make the symmetric structure of the expression explicit, while keeping the argument of the logarithm dimensionless.}
\begin{align}
I(m_1,m_2,m_3,m_4)&= -\frac{1}{16\pi ^2} 
\sum_{i} \frac{{m_i}^2 \log
   \left(\frac{{m_i}^2}{\mu^2}\right)}{\Pi_{k\neq i}\left({m_i}^2-{m_k}^2\right)}\;.
\end{align}
Evaluation of the sums leads to the following flavor structure of the neutrino mass matrix:
\begin{align}\label{eq:NuMass}
 {M_\nu}=\left(\begin{array}{ccc}
\hat{a} &\hat{e}\,e^{\I\alpha_{\lambda}}&\hat{e}\,e^{\I\alpha_{\lambda}}\\
.&\hat{a}+\hat{b}\,e^{\I\alpha_{\lambda}}& \hat{d}+\hat{e}\,e^{\I\alpha_{\lambda}}\\
.&.&\hat{a}
\end{array}\right)\;,
\end{align}
where the four real coefficients are given by
\begin{subequations}\label{eq:neutrino-mass-expressions}
\begin{align}
 \hat{a}&=\frac{1}{36\sqrt{3}}\,h_1^2\, \lambda_3\, \lambda_{1}  v^2 \, ( a c+b d)\, M_{S}\, I\left(M_1,M_1,M_3,M_S\right),\\
 \hat{d}&=\frac{1}{72\sqrt{3}}\, h_1 h_2\, \lambda_4\,  \lambda_1
   v^2\, ( b c- a d)\, M_{S}\, I(M_1,M_2,M_3,M_S),\\
 \hat{b}&=\frac{1}{108}\, h_2^2\, \lambda_4\,   \lambda_2 v^2 \,   ( b c -a d)\, M_{S} \,I(M_2,M_2,M_3,M_S),\\
 \hat{e}&=\frac{1}{216}\, h_1 h_2 \, \lambda_3\, \lambda_2 
   v^2\,  ( a c+b d)\,M_{S}\, I(M_1,M_2,M_3,M_S).
\end{align}
\end{subequations}
Hence,  neutrino masses are suppressed by one insertion of the EW breaking VEV $\lambda_1\ev{\chi^2}/M_0^2$, with $M_0$ being the largest mass of the particles in the loop $M_0\sim\max_{i=1,2,3,S}M_i$, and one mass insertion of the flavor breaking VEV $\lambda_2\ev{\phi_1\phi_2}/M_0^2$.
A phenomenologically viable neutrino mass scale is obtained for e.g. $M_{0}\sim\Order{\TeV}$, $\ev{\chi},\ev{\phi_i}\sim\Order{100\GeV}$ and $h_i,\lambda_i\sim\Order{0.01-0.1}$. 
The next-to-leading order corrections are suppressed by $\lambda_1\ev{\chi^2}/M_0^2$ or $\lambda_2\ev{\phi_1\phi_2}/M_0^2$, which amounts to an $\Order{0.0001-0.001}$ correction for our typical values and can be neglected to a good approximation.

The neutrino mass elements correspond to the following operators:
\begin{itemize}
\item $\hat{a}:\left(L^T \sigma_2 \vec{\sigma} L\right)_{\MoreRep{1}{1}} \left(\chi^T \tau_2 \vec{\tau} \chi\right)_{\MoreRep{1}{1}} \left(\phi_1 \phi_2\right)_{\MoreRep{1}{1}}$
\item $\hat{d}:\left(L^T \sigma_2 \vec{\sigma} L\right)_{\MoreRep{3}{1}} \left(\chi^T \sigma_2 \vec{\sigma} \chi\right)_{\MoreRep{1}{1}} \left(\phi_1 \phi_2\right)_{\MoreRep{3}{1}}$
\item $\hat{e}:\left(L^T \sigma_2 \vec{\sigma} L\right)_{\MoreRep{3}{1}} \left(\chi^T \sigma_2 \vec{\sigma} \chi\right)_{\MoreRep{3}{1}} \left(\phi_1 \phi_2\right)_{\MoreRep{1}{1}}$
\item $\hat{b}:\sum_i \omega ^ {i-1} \left(L^T \sigma_2 \vec{\sigma} L\right)_{\MoreRep{1}{i}} \left[\left(\chi^T \sigma_2 \vec{\sigma} \chi\right)_{\MoreRep{3}{1}} \left(\phi_1 \phi_2\right)_{\MoreRep{3}{1}}\right]_{\MoreRep{1}{i}^*}$
\end{itemize}
The fact that only the combination shown in the last line contributes to neutrino masses is due to the UV completion presented here. In a general theory one might have all operators present, thereby reducing the predictability of the theory.

\addtocontents{toc}{\protect\setcounter{tocdepth}{1}}
\subsection{Phenomenological Implications}
\addtocontents{toc}{\protect\setcounter{tocdepth}{2}}
As the neutrino mass matrix is described by five physical real parameters, there are four predictions in the lepton sector at leading order. They can easily be read off from \Eqref{eq:NuMass} in terms of matrix elements, but the expressions in terms of mixing parameters are non-trivial.
In the flavor basis, where the charged lepton mass matrix is diagonal, the neutrino mass matrix is given by
\begin{equation}\label{eq:NuMassFl}
{M_\nu}^{fl}=\left(\begin{array}{ccc}
\hat a + \frac{2 \hat d}{3} + \left(2\hat e+\frac{\hat b}{3}\right)\, e^{\I\alpha_\lambda}
 & -\frac{\hat d}{3} +\frac{\hat b}{3} e^{\I\alpha_\lambda}\omega^2 
 &-\frac{\hat d}{3} +\frac{\hat b}{3} e^{\I\alpha_\lambda}\omega \\
. & \frac{2\hat d}{3}+\frac{\hat b}{3} e^{\I\alpha_\lambda} \omega
& \hat a -\frac{\hat d}{3}+\left(\frac{\hat b}{3}-\hat e\right) e^{\I\alpha_\lambda}\\
.&.& \frac{2\hat d}{3} +\frac{\hat b}{3} e^{\I\alpha_\lambda}\omega^2\\
\end{array}\right)\;.
\end{equation}
and it is instructive to look at the neutrino mass matrix in the tri-bimaximal basis ${M_\nu}^{tbm}=U_{HPS}^T {M_\nu}^{fl} U_{HPS}$, i.e., \begin{equation}
{M_\nu}^{tbm}=
\left(
\begin{array}{ccc}
 \hat a+\hat d+\left(\frac{\hat b}{2}+ \hat e\right)\,e^{\I \alpha_\lambda } & -\sqrt{2}\,\hat e\, e^{\I \alpha_\lambda }  & -\I\frac{\hat b}{2} \,  e^{\I \alpha_\lambda} \\
. & \hat a & 0 \\
 .&. & -\hat a+\hat d+\left(\hat e-\frac{\hat b}{2}\right)\,  e^{\I \alpha_\lambda}
\end{array}
\right)\;.
\label{eq:mass-matrices-in-TBM-basis}
\end{equation}
We will first discuss limiting cases analytically and then perform a numerical analysis of the general neutrino mass matrix.
In the limit of $|\lambda_{2}|  v^2 \rightarrow0$, both $\hat b$ and $ \hat e$ vanish and we obtain tri-bimaximal mixing 
\begin{align}
U_{HPS}\equiv\Omega_T^\dagger \Omega_U=\left(
\begin{array}{ccc}
\sqrt{\frac{2}{3}}&\frac{1}{\sqrt{3}}&0\\
-\frac{1}{\sqrt{6}}&\frac{1}{\sqrt{3}}&\frac{1}{\sqrt{2}}\\
-\frac{1}{\sqrt{6}}&\frac{1}{\sqrt{3}}&-\frac{1}{\sqrt{2}} \end{array}\right)
\qquad\quad\mathrm{with}\qquad
\Omega_U=\left(
\begin{array}{ccc}
 0 & 1 & 0 \\
 \frac{1}{\sqrt{2}} & 0 & -\frac{i}{\sqrt{2}} \\
 \frac{1}{\sqrt{2}} & 0 & \frac{i}{\sqrt{2}}
\end{array}
\right)\;.
\label{eq:HPS-matrix}
\end{align} 
From Eq.~\eqref{eq:mass-matrices-in-TBM-basis} we can read off that switching on $\hat e\neq 0$ while keeping $\hat b= 0$ results in a correction to the PMNS matrix of the form 
$U=U_{HPS}U_{12}(\tilde \theta_{12}) P$ with $U_{12}(\tilde\theta_{12})$ denoting the unitary matrix 
\begin{equation}
U_{12}(\tilde \theta)=\left(\begin{array}{ccccc}
c_{12}&  -s_{12} e^{-\I\delta_{12}} &\\
s_{12} e^{\I\delta_{12}}&c_{12} &\\
&&1\\
\end{array}\right),\label{eq:def-12}
\end{equation}
with $c_{12}=\cos\tilde \theta_{12}$, $s_{12}=\sin\tilde\theta_{12}$
and $P$ being an arbitrary phase matrix. In the standard parameterization of the PMNS matrix~\cite{PDG:2012} with the 1-2 rotation to the right, this 1-2 correction only affects the solar angle, while maintaining the predictions of a maximal atmospheric and vanishing reactor angle. Since large corrections to this angle are not allowed, in the phenomenologically acceptable parameter space the relations $\hat e \ll {\hat b,\hat a,\hat d} $ should hold. 

\begin{figure}\centering
\includegraphics[width=.75\textwidth] {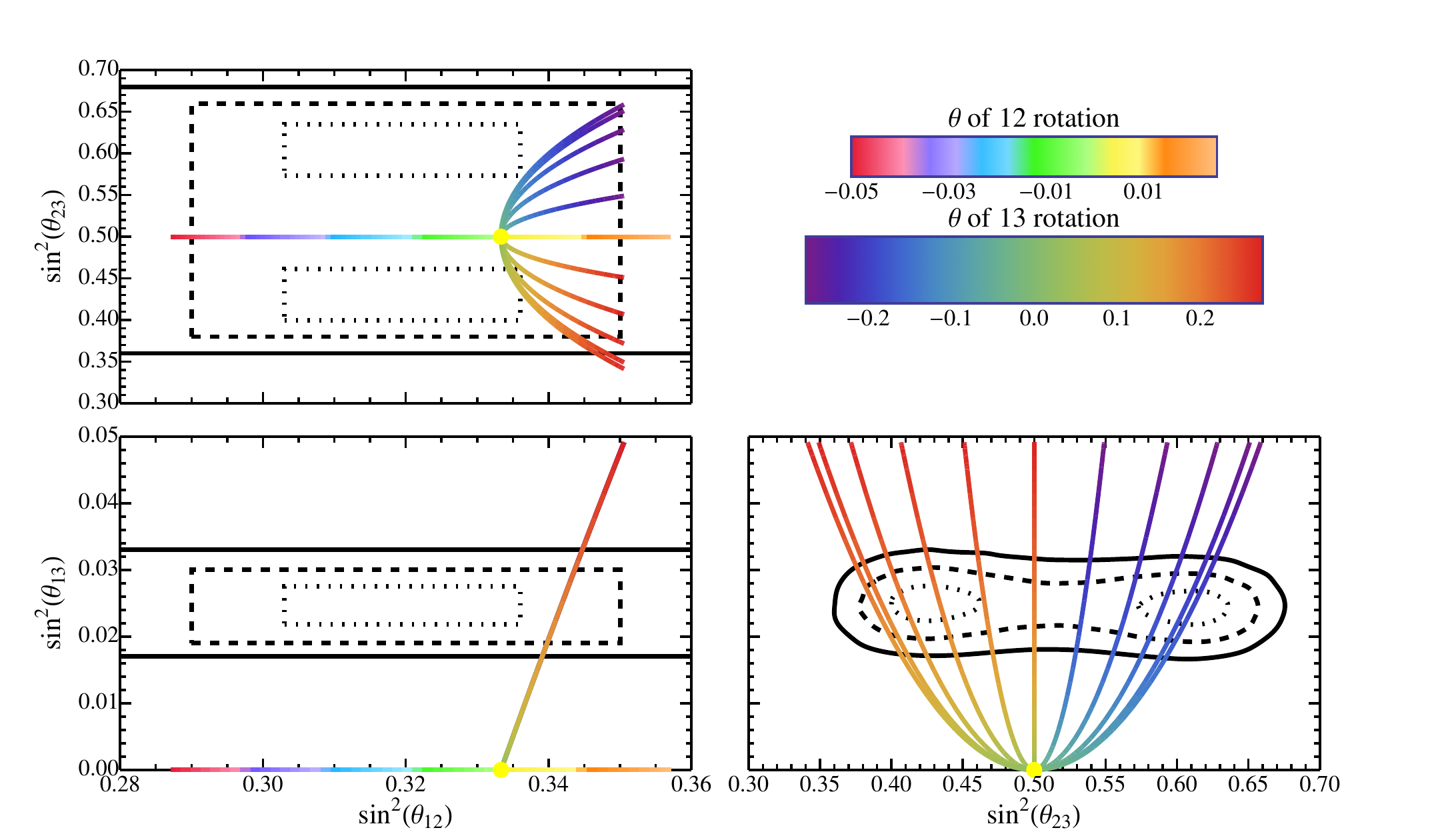} 
\caption[]{The deviations from tri-bimaximal mixing of the form $U=U_{HPS}U_{12}$ and $U=U_{HPS}U_{13}$ generated by the angle $\theta$ defined in \Eqref{eq:def-12}. The yellow point represents TBM, the continuous lines give the deviations from TBM with the angle $\theta$ given by the color codes in the top right corner for $\delta=\frac{n}{5}\frac{\pi}{2}$ for $n=0,\dots,5$, where $n=0$ is the outermost parabola etc. The one, two and three sigma regions of a recent global fit~\cite{Tortola:2012te} are indicated by dotted, dashed and continuous contours, respectively.\label{fig:12-and-13-plot}}
\end{figure}

On the other hand, if we take $\hat b\neq 0$ while $\hat e= 0$, we see from Eq.~\eqref{eq:mass-matrices-in-TBM-basis} that this requires a 1-3 correction $U=U_{HPS}U_{13}(\tilde \theta_{13})P$, where $U_{13}(\tilde \theta_{13})$, analogous to $U_{12}(\tilde \theta_{12})$, denotes a complex rotation in the 1--3 plane. This correction is of the trimaximal mixing~\cite{Haba:2006dz,He:2006qd,Grimus:2008tt,Ishimori:2010fs,He:2011gb,Antusch:2011ic,Cooper:2012wf,King:2011zj} form, which can perturb TBM back into agreement with experiment. The effect of the various deviations from TBM is illustrated in \Figref{fig:12-and-13-plot}. 

To gain an analytical understanding of how the additional parameters affect the mixing angles, we can perform a perturbative analysis in the limit of small $\hat e$ and therefore small $|\sin^2\theta_{12}-\frac13|$. The PMNS matrix can be described by $U_{HPS}U_{13}(\tilde\theta_{13})U_{12}(\tilde\theta_{12})P$, where $\tilde\theta_{12}$ and $\tilde\theta_{13}$ are small in the phenomenologically interesting region and the Majorana phases are given by $P=\diag( e^{\I \alpha_1/2}, 1,e^{\I \alpha_3/2})$. Hence, we can permute the matrices $U_{12}$ and $U_{13}$ and we define $r_{1i}=\sin\tilde \theta_{1i} \cos\tilde \delta_{1i}$ and $t_{1i}=\tan\tilde \delta_{1i}$, which evaluate to
\begin{subequations}
\begin{align}
r_{13}&=\frac{\hat b \sin \alpha_\lambda}{4 \hat a+2 \hat b \cos \alpha_\lambda }\;,&
t_{13}&=\frac{2 \hat a \cos \alpha_\lambda +\hat b}{2 \hat d\sin\alpha_\lambda}\;,\\
r_{12}&=\frac{\sqrt{2} \hat d \hat e \sin \alpha_\lambda }{\Delta m_{21,0}^2}\;,&
t_{12}&=\frac{ 2 (2 \hat a+\hat d) \cos \alpha_\lambda +\hat b}{2 \hat d\sin \alpha_\lambda }\;,
\end{align}
\end{subequations}
where $\Delta m_{21,0}^2$ is the leading order solar mass squared difference, i.e. neglecting the small corrections of $r_{13}$ and $\hat e$.
The phases of the matrix $P$ are given by
\begin{subequations}
\begin{align}
\tan\alpha_1&=\frac{2 \hat b\, r_{13} \cos \alpha_\lambda -\sin \alpha_\lambda  (2 \hat b\, r_{13} t_{13}+\hat b+2 \hat e)}{2
   (\hat a+\hat b\, r_{13} \sin \alpha_\lambda +\hat d)+\cos \alpha_\lambda  (2 \hat b\, r_{13} t_{13}+\hat b+2
   \hat e)},\\
\tan\alpha_2&=\frac{\sin \alpha_\lambda  (\hat b (2 r_{13} t_{13}-1)+2 \hat e)+2 \hat b\, r_{13} \cos \alpha_\lambda }{2
   (\hat a+\hat b\, r_{13} \sin \alpha_\lambda -\hat d)+\cos \alpha_\lambda  (-2 \hat b\, r_{13} t_{13}+\hat b-2
   \hat e)}\;.
\end{align}
\end{subequations}
Similar to~\cite{King:2007pr}, we can parameterize the leptonic mixing matrix 
in terms of deviations from the tri-bimaximal mixing angles as defined in \Eqref{eq:devTBM}.
The Dirac CP phase $\delta_{CP}$ is undefined in the tri-bimaximal mixing limit and we leave it free and do not expand in it. Besides the contributions of $\alpha_{1,3}$ to the Majorana phases $\varphi_{1,2}$ in the standard parameterization, there are also small corrections $\delta\varphi_{1,2}$ from the matrices $U_{12}(\tilde\theta_{12})$ and $U_{13}(\tilde\theta_{13})$ resulting in
\begin{align}
\varphi_{1}&=\alpha_1-\alpha_3+\delta\varphi_1,& &\mathrm{and}&
\varphi_2&=\pi-\alpha_3+\delta\varphi_2\;.
\end{align}
This expansion leads to the following form of the PMNS matrix
\begin{equation}
U=\left(
\begin{array}{ccc}
 \frac{s+i \delta \varphi _1-2}{\sqrt{6}} & \frac{2 i (s+1)+\delta \varphi _2}{2 \sqrt{3}} & -\frac{e^{-i \delta } r}{\sqrt{2}} \\
 \frac{2 \left(-a+e^{i \delta } r+s+1\right)-i \delta \varphi _1}{2 \sqrt{6}} & \frac{\delta \varphi _2-i \left(2 a+e^{i \delta }
   r+s-2\right)}{2 \sqrt{3}} & -\frac{a+1}{\sqrt{2}} \\
 \frac{2 \left(a-e^{i \delta } r+s+1\right)-i \delta \varphi _1}{2 \sqrt{6}} & \frac{i \left(2 a+e^{i \delta } r-s+2\right)+\delta \varphi
   _2}{2 \sqrt{3}} & -\frac{a-1}{\sqrt{2}}
\end{array}
\right)\,P\;.
\end{equation}
Equating the expanded form of $U$ to $U_{HPS}U_{13}(\tilde\theta_{13})U_{12}(\tilde\theta_{12})P$ determines all free parameters $s,r,a,\delta,\delta\varphi_1,\delta\varphi_2$ as well as some corrections to unphysical phases, which we suppressed for simplicity. The
first order deviations from the mixing angles are 
\begin{align}
s&=-\sqrt{2} r_{12} t_{12},&
r\cos\delta&=-\frac{2 r_{13}  }{\sqrt{3}},&
a&=\frac{r_{13}}{\sqrt{3}},
\end{align}
and the CP phases  are given by
\begin{align}
\tan\delta_{CP}&= \tan\tilde\delta _{13}&
\varphi_1&=\alpha_1-\alpha_3-2\sqrt{2} r_{12}&
\varphi_2&=\pi-\alpha_3-2\sqrt{2} r_{12}\;.
\end{align}
Following~\cite{King:2007pr}, we can derive a sum rule, which relates the deviations of the atmospheric mixing angle with the ones of the reactor mixing angle
\begin{equation}
a=-\frac12 r\cos\delta_{CP}\;.
\label{eq:atmos-sum-rule}
\end{equation}
\begin{figure}[tb]
\centering
\includegraphics[width=1\textwidth]{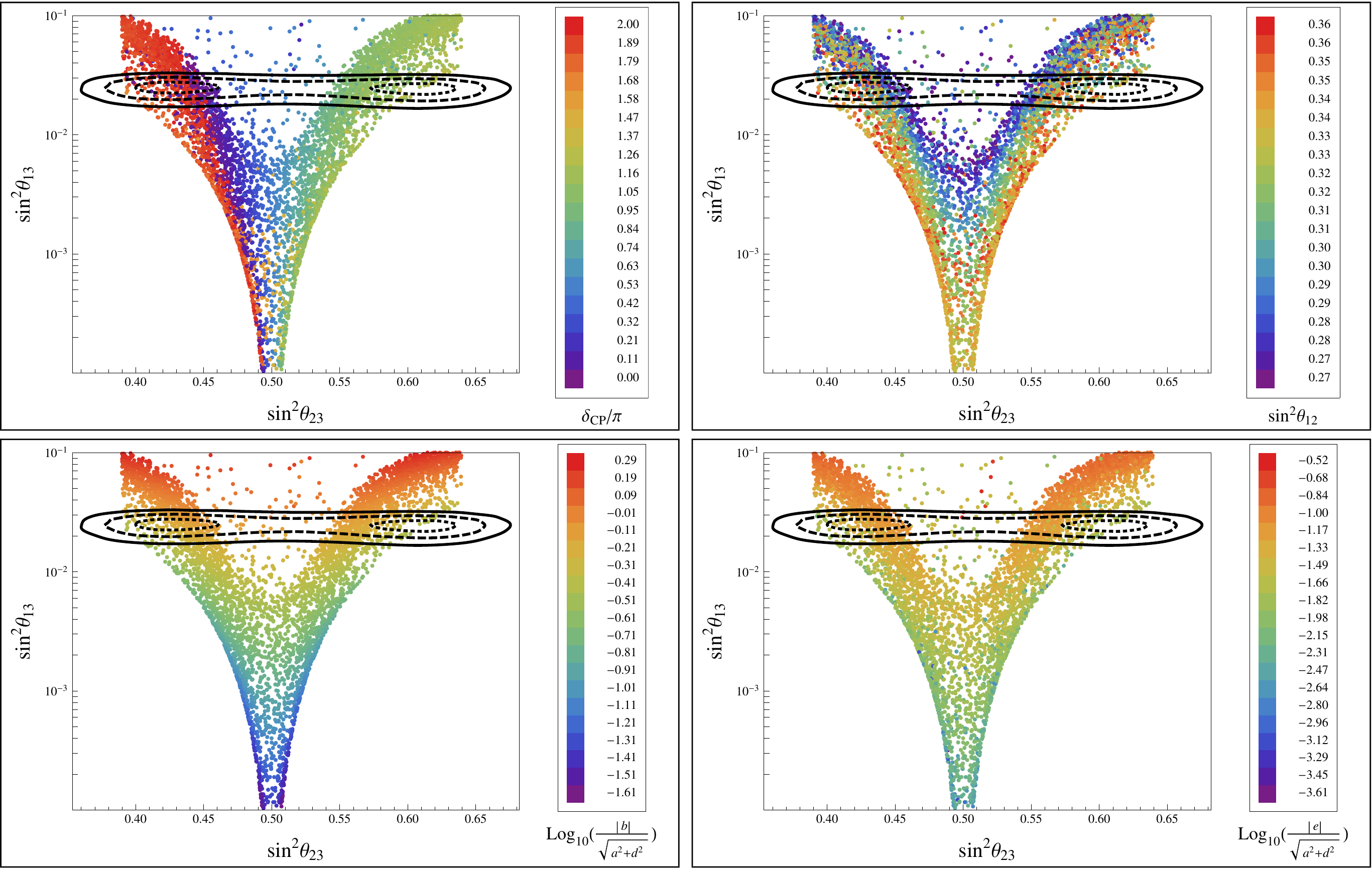}
\caption{Dependence of the reactor angle $\theta_{13}$ on the atmospheric mixing angle $\theta_{23}$. The various color codings are given next to each scatter plot. Top left: For $\sin^2 \theta_{23}<1/2$( $\sin^2 \theta_{23}>1/2$ ) the model predicts $ \delta_{CP}=0,2 \pi$($ \delta_{CP}=\pi$). Top right: The scatterplot shows a band structure in $\sin^2 \theta_{12}$. Bottom left: For the points in the experimentally allowed region, $\hat b$ has to be of similar size as $\hat a, \hat d$. Bottom right: For the points in the experimentally allowed region, $\hat e$ has to be of approximately one order of magnitude smaller than $\hat a, \hat d$. The one, two and three sigma regions of Ref.~\cite{Tortola:2012te} are again indicated by dotted, dashed and continuous contours, respectively.
\label{fig:angle}}
\end{figure}
The masses are determined by
\begin{align}
m_1^2&=\hat a^2+\hat b\, (\hat a+\hat d) \cos \alpha_\lambda+2 \hat a \hat d+\frac{\hat b^2}{4}+\hat d^2\;,&
m_2^2&=\hat a^2\;,\\\nonumber
m_3^2&=\hat a^2+\hat b\, (\hat a-\hat d) \cos \alpha_\lambda-2 \hat a \hat d+\frac{\hat b^2}{4}+\hat d^2,
\end{align}
to leading order in the small mixings $r_{13}$, $r_{12}$, and the leading order ratio of mass squared differences is given by
\begin{align}
\frac{\Delta m_{21}^2}{\Delta m_{32}^2}&=\frac{4 \hat a\,(2\hat d+\hat b\cos\alpha_\lambda)+4\hat d\, (\hat d+\hat b \cos\alpha_\lambda)+\hat b^2}
{4 \hat a\,(2\hat d-\hat b\cos\alpha_\lambda)-4\hat d\, (\hat d-\hat b \cos\alpha_\lambda)-\hat b^2}\;.
\end{align}
At next-to leading order, $m_1$ and $m_3$ receive corrections
\begin{subequations}
\begin{align}
\delta m_1^2&=\hat b (2 r_{13} (\hat a+\hat d) \sin \alpha_\lambda+\hat b\, r_{13} t_{13}+\hat e)+2 (\hat a+\hat d)\, \cos \alpha_\lambda (\hat b\, r_{13} t_{13}+\hat e)\;,\\
\delta m_3^2&=-\hat b (-2 r_{13} (\hat a-\hat d) \sin \alpha_\lambda+\hat b\, r_{13} t_{13}+\hat e)-2
   (\hat a-\hat d) \cos \alpha_\lambda (\hat b\, r_{13} t_{13}+\hat e)\;.
\end{align}
\end{subequations}
\begin{figure}
\begin{minipage}{0.48\linewidth}
\includegraphics[width=\textwidth]{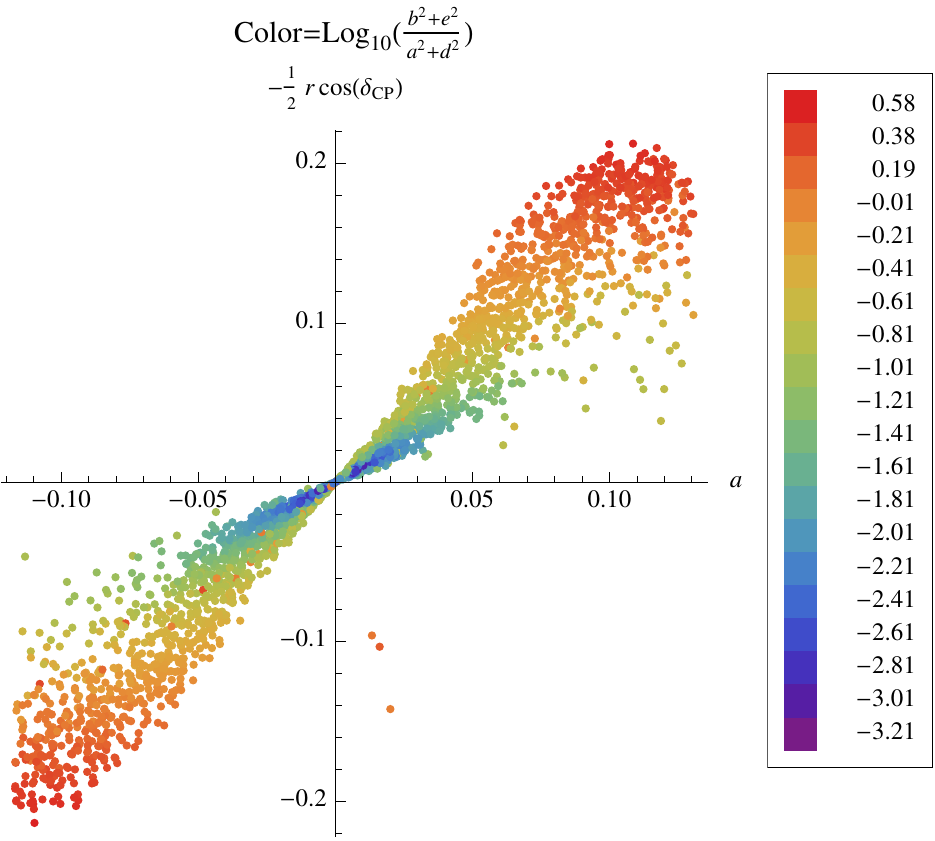}
   \caption{ Numerical evaluation of the approximate atmospheric sum rule \eqref{eq:atmos-sum-rule}. The numerical evaluation shows that the sum rule holds to a good degree of approximation.\label{fig:r/a}}
\end{minipage}\hfill
\begin{minipage}{0.48\linewidth}
\includegraphics[width=\textwidth]{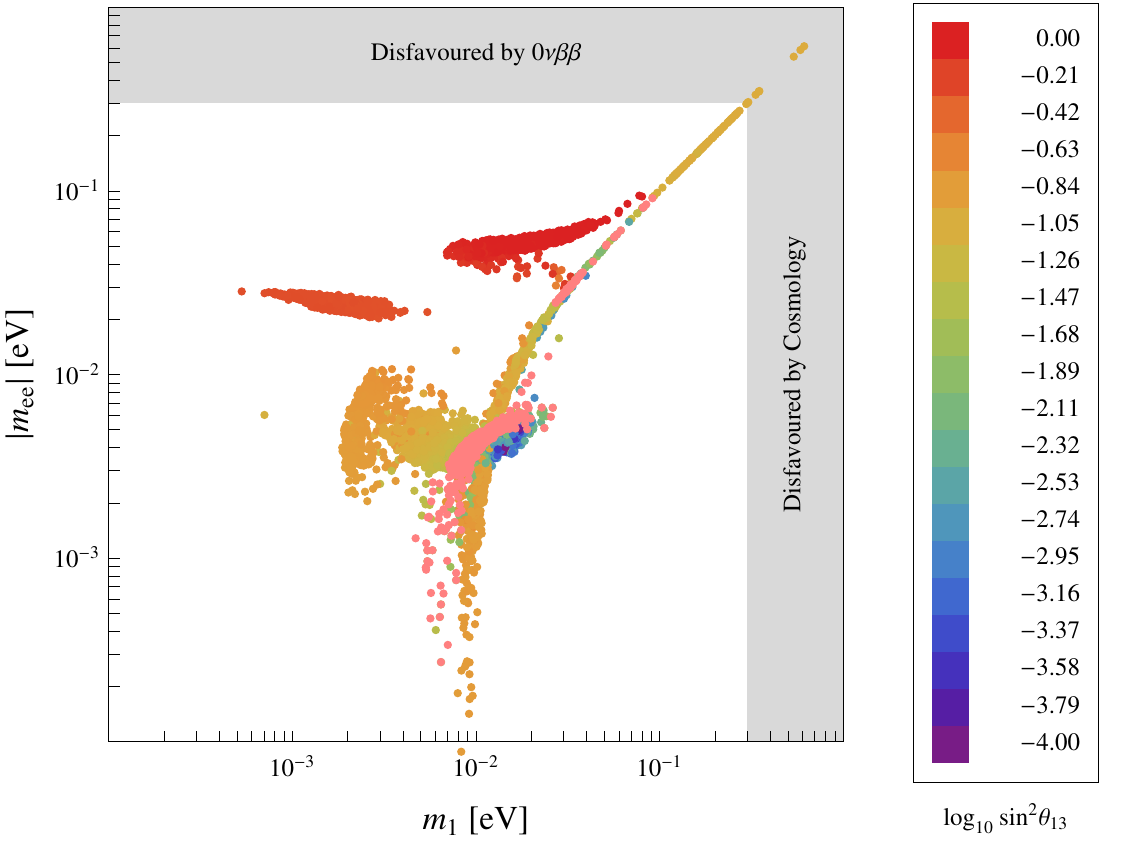}
\caption{Expectation for the effective mass of neutrinoless double beta decay. The pink points lie within the 3 sigma region for all oscillation parameters. The points with color coding lie within the 3 sigma range for all observables except $\theta_{13}$. \label{fig:0nu2beta}}   
\end{minipage}
\end{figure}
To illustrate our findings numerically, we have performed a numerical scan over the model's parameter space. We have randomly drawn values for the model parameters of order unity, assuming a Gaussian distribution with an expectation value of one and a variance of $0.5$.
The plots in \Figref{fig:angle} show the relation between the atmospheric mixing angle $\theta_{23}$ and the reactor angle $\theta_{13}$. From the bottom two plots one can read off that $\hat b$ is of the same order as $\hat a$ and $\hat d$ for the experimentally measured $\theta_{13}$ while $\hat e$ has to be about one order of magnitude smaller. The color codings of the two top panels show the mixing parameters $\delta_{CP}$ and $\sin^2 \theta_{12}$. Clearly, the model is predictive: if $\sin^2 \theta_{23}$ is found to be close to the best fit point in the octant with $\sin^2 \theta_{23}<1/2$, the prediction for the CP phase is $ \delta_{CP}=0,2 \pi$ while for $\sin^2 \theta_{23}>1/2$ it is predicted to be $ \delta_{CP}=\pi$. To establish the correlation with $\sin^2 \theta_{12}$ shown in the top right panel, a precision determination of all the mixing angles is needed. In \Figref{fig:r/a}, as a consistency check of our analytical expressions, the atmospheric sum rule \eqref{eq:atmos-sum-rule} is shown for the points obtained in the numerical scan. The color coding gives an indication of the magnitude of deviations from TBM and for small values the approximate relation is fulfilled to good accuracy.

Finally, let us comment on the predictions for neutrinoless double beta decay. 
As can be read off from \Eqref{eq:NuMassFl}, the effective Majorana mass of the electron neutrino is given by
\begin{equation}
|m_{ee}|=\left|\hat a + \frac{2 \hat d}{3} + \left(2\hat e+\frac{\hat b}{3}\right)\, e^{\I\alpha_\lambda}\right|\;,
\end{equation}
which can be expressed in terms of physical parameters as
\begin{equation}
|m_{ee}|=\left|\sum_i U_{ei}^2 m_i\right|\approx \frac{2m_1-m_2}{3}\left|1-\frac{2m_1+2m_2}{2m_1-m_2}s-\I\frac{2\delta\varphi_1 m_1-\delta\varphi_2 m_2}{2m_1-m_2}\right|\;.
\end{equation}
As the additional neutral fermions $S$ do not mix with neutrinos, there is no additional contribution due to the heavy singlet, like in Ma's scotogenic model~\cite{Ma:2006km,Kubo:2006kx}. In \Figref{fig:0nu2beta} we show the predicted range for the effective Majorana mass of the electron neutrino. As can be seen, the scan of parameters prefers moderately large values of the absolute mass scale, however, the effective Majorana mass of the electron neutrino can become small or even vanish.

%%%%%%%%%%%%%%%%%%%%%%%%%%%%%%%%%%%%%%%%%%%%%%%%%%%%%%%%%%%%%%%%%%%55
%%%%%%%%%%%%%%%%%%%%%%%%%%%%%%%%%%%%%%%%%%%%%%%%%%%%%%%%%%%%%%%%%%%55
%%%%%%%%%%%%%%%%%%%%%%%%%%%%%%%%%%%%%%%%%%%%%%%%%%%%%%%%%%%%%%%%%%%55
%\input{lfv-v1}
\section{Lepton Flavor Violation}
\label{sec:LFVprocesses}

In models with radiative neutrino mass generation, generally the particles in the loop can also mediate flavor changing processes, 
in particular lepton flavor violating rare decays. Before we enter into a detailed discussion of the various processes, we want to remind the reader about the remnant $Z_3$ symmetry in the charged lepton sector 
\begin{align}
\left(H, {\varphi^{\prime}}, {\varphi^{\prime \prime}} \right)\sim (1,\omega^2,\omega), \qquad \left(L_e, L_\mu, L_\tau \right)\sim (1,\omega^2,\omega),\qquad
(e^c,\mu^c,\tau^c)\sim (1,\omega,\omega^2),
\end{align}
which suppresses several LFV rare decays. If the remnant $Z_3$ would be a symmetry of the whole Lagrangian, only the following LFV rare decays
\begin{equation*}
\tau^+\rightarrow \mu^+\mu^+e^- 
\quad\quad\text{and}\quad\quad
\tau^+\rightarrow e^+e^+\mu^- 
\end{equation*}
and their charged conjugates would be allowed. All other decays can only proceed through a coupling to the $Z_3$ breaking VEVs of the neutrino sector. Those decays are naturally suppressed and the symmetry thus protects the model from large constraints. At first, we will discuss the radiative LFV rare decays $l_i\rightarrow l_j\gamma$ in \Secref{sec:radLFV}, focusing on the experimentally most well studied process, namely the process
$\mu\rightarrow e \gamma$. In \Secref{sec:LFVintoLeptons}, we discuss the LFV rare decays with purely leptonic final states, which are allowed at tree level, but suppressed by a three-body final state. Finally, we calculate the anomalous magnetic moment of the muon and compare it to experiment in \Secref{sec:anomMagMom}.

\addtocontents{toc}{\protect\setcounter{tocdepth}{1}}
\subsection{Radiative LFV Decays $l_i\rightarrow l_j \gamma$}
\addtocontents{toc}{\protect\setcounter{tocdepth}{2}}

%\mathversion{bold}
%\subsection*{Radiative LFV Decays $l_i\rightarrow l_j \gamma$}
%\mathversion{normal}
\label{sec:radLFV}
Let us first discuss the process of type $l_i\rightarrow l_j \gamma$ using an effective field theory approach. Such processes are described by effective operators of the form~\cite{Feruglio:2008ht,Borzumati:1986qx}
\begin{equation}\label{eq:effOpRadLFV}
L  \sigma_{\mu \nu} F^{\mu \nu}  \ell^c \tilde{H}/M^2 \sim \left(\MoreRep{3}{1}, 1 \right),
\end{equation}
which transforms in the same way as the mass term under the flavor symmetry. It thus has to be multiplied by flavons to form an invariant. As we already mentioned, the remnant $Z_3$ symmetry in the charged lepton sector forbids all radiative LFV rare decays. Hence, the effective operator in \Eqref{eq:effOpRadLFV} has to involve VEVs of the neutrino sector in order to lead to non-vanishing decay rates.
The lowest order operators that can multiply the mentioned LFV operator in the flavor basis read
\begin{subequations}
\begin{align}
 \Omega_T^\dagger\vev{\left(\phi_1^4\right)_{\MoreRep{3}{1}}}&=\frac{1}{6}\left(ab(b^2-a^2)\right)(1,1,1)^T,\\
 \Omega_T^\dagger\vev{\left(\phi_2^4\right)_{\MoreRep{3}{1}}}&=\frac{1}{6}\left(cd (d^2-c^2) \right)(1,1,1)^T,\\
 \Omega_T^\dagger\vev{\left(\phi_1^2 \phi_2^2\right)_{\MoreRep{3}{1}}}&=\frac{1}{3}\left(a b (c^2-d^2) \right)(1,1,1)^T,\\
 \Omega_T^\dagger\vev{\left(\phi_1^2 \phi_2^2\right)_{\MoreRep{3}{1}}}&=\frac{1}{3}\left(c d (a^2-b^2) \right)(1,1,1)^T.
\end{align}
\end{subequations}
There can be more than one contraction, but in the vacuum they all result in these expressions. The lowest order effective operators thus all give contributions that can be written as
\begin{align}
\mathcal{L}_{eff}=\I \frac{e}{M^2} {\ell^c}^T H^\dagger \sigma_{\mu \nu} F^{\mu \nu} \mathcal{M} L+\hc \qquad
\mathrm{with}\qquad \mathcal{M}=\left(\begin{array}{ccc}\alpha_1&\alpha_1&\alpha_1\\ \alpha_2&\alpha_2&\alpha_2\\\alpha_3&\alpha_3&\alpha_3\end{array} \right) \frac{\vev{\phi_1^4}}{M^4}
\end{align}
where $\alpha_i$ are dimensionless couplings that should (naturally) be of order one and the mass scale M is the suppression scale of the higher dimensional operators. Note that the structure of flavor symmetry breaking in the neutrino sector is encoded in $\mathcal{M}$. The symmetry thus automatically leads to a large suppression. From this matrix the LFV transition amplitudes can be determined as~\cite{Feruglio:2008ht}
\begin{align}
\frac{\mathrm{Br}(l_i\rightarrow l_j \gamma) }{\mathrm{Br}(l_i\rightarrow l_j \nu_i \overline{\nu}_j)}=\frac{12 \sqrt{2} \pi^3 \alpha }{G_F^3 m_i^2 M^4}\left( \abs{\mathcal{M}_{ij}}^2+\abs{\mathcal{M}_{ij}}^2\right)^2
\end{align}
and the magnetic dipole moments $a_i$ and electric dipole moments $d_i$ of the charged leptons are given by~\cite{Feruglio:2008ht}
\begin{align}
a_i=2 m_i \frac{v}{\sqrt{2} M^2} \re \mathcal{M}_{ii}, \qquad d_i=e \frac{v}{\sqrt{2} M^2} \im \mathcal{M}_{ii}.
\end{align} 
Note that the matrix $\mathcal{M}$ has additional dominant contributions to the diagonal entries stemming from operators that involve $\chi$ instead of $(\phi_i)^4$. Using only the observables $\mu\to e\gamma$, $\tau\to \mu\gamma$ and $\tau\to e\gamma$ as well as charged lepton electric and dipole moments, it is therefore very difficult to test the underlying symmetry pattern, but it can give important indications distinguishing different models. For example in this model one would expect -- barring the possibility of fine-tuned cancellations among the $\alpha_i$ -- similar branching ratios for the LFV decays $\mu\to e\gamma$, $\tau\to \mu\gamma$ and $\tau\to e\gamma$, as was also found in SUSY $A_4$ models~\cite{Feruglio:2008ht,Feruglio:2009hu}.

\begin{figure}
\begin{center}
\begin{subfigure}[b]{0.3\textwidth}
\includegraphics[width=\textwidth]{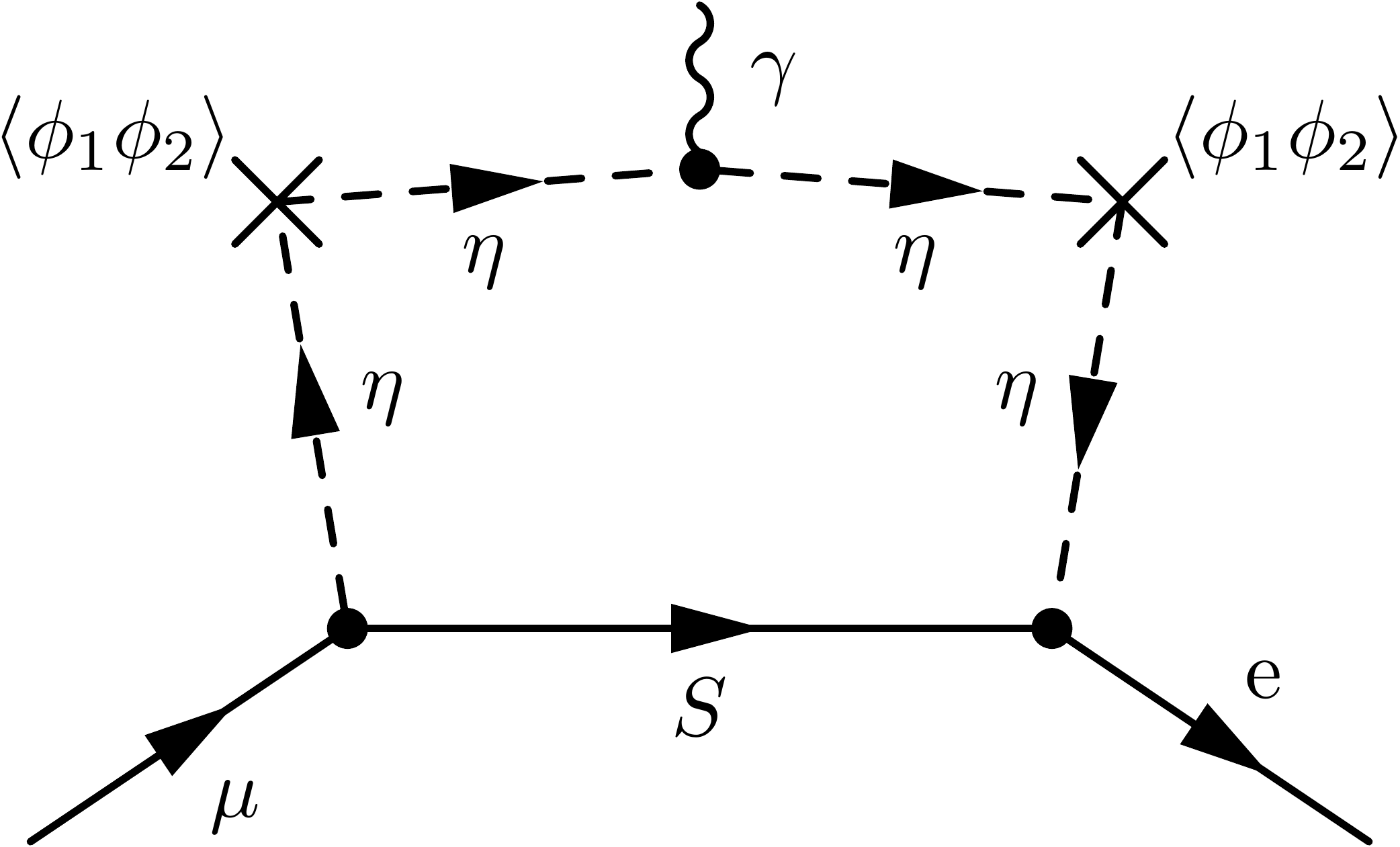}
\caption{Lowest order $\mu\rightarrow e \gamma$  process mediated by $\eta$'s.\label{fig:LFV-mu-to-e-gamma}}
\end{subfigure}\qquad\qquad\qquad
\begin{subfigure}[b]{0.3\textwidth}
\includegraphics[width=\textwidth]{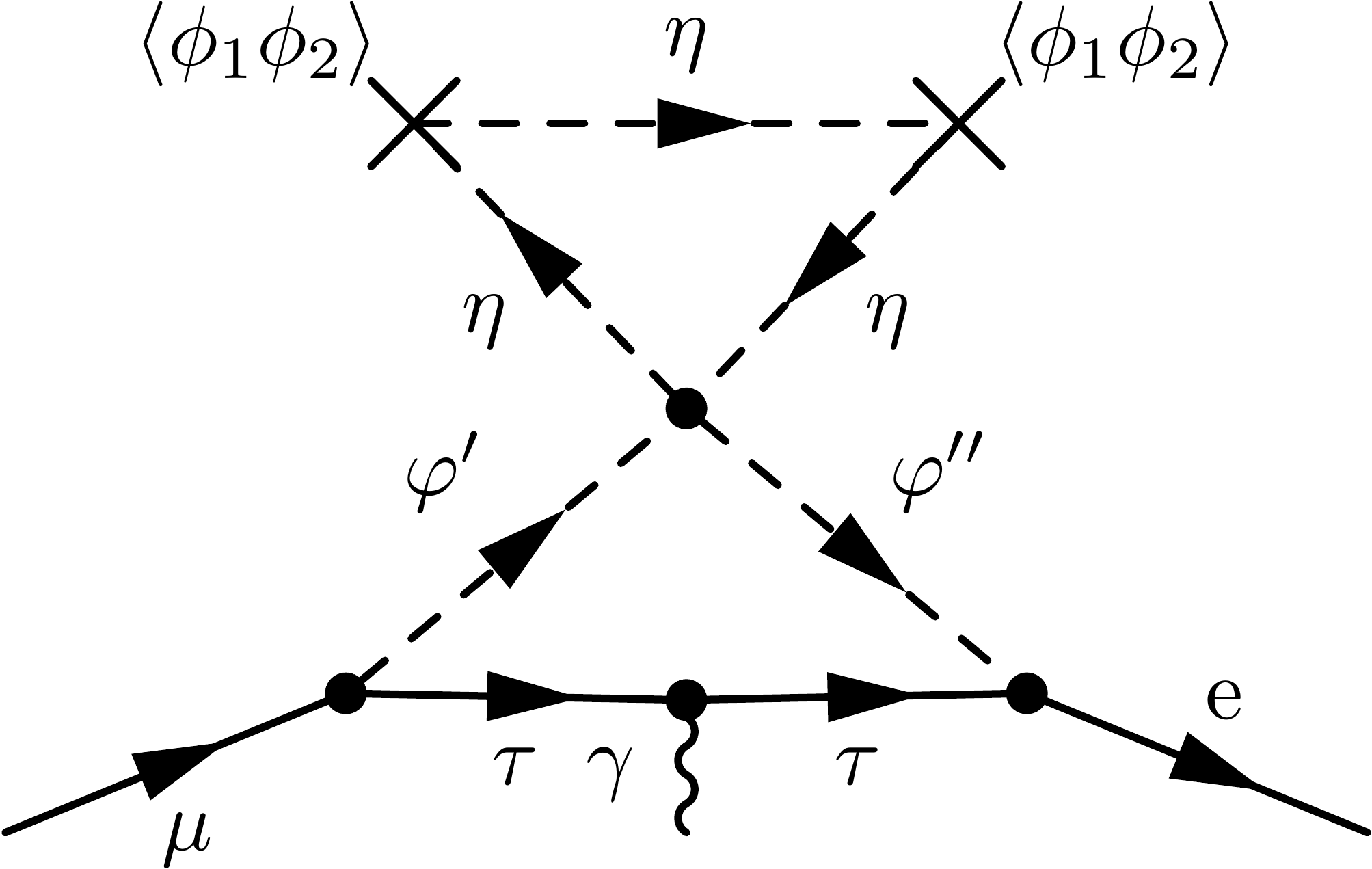}
\caption{Lowest order $\mu\rightarrow e \gamma$  process mediated by ${\varphi^{\prime}}$.\label{fig:LFV-mu-to-e-gamma-2loop}}
\end{subfigure}
\end{center}
\caption{Lowest order $\mu\rightarrow e \gamma$ processes mediated by $\hat\eta$ (left) and $\varphi^{\prime (\prime)}$ (right). There has to a coupling to the VEVs $\vev{\phi_1 \phi_2}$ of the neutrino sector, which suppresses the amplitudes.}
\end{figure}

In the following, we will focus on $\mu\to e\gamma$, which is the most tightly constrained LFV rare decay.
The leading contribution to $\mu\rightarrow e\gamma$ is given by the diagram depicted in \Figref{fig:LFV-mu-to-e-gamma}. It is similar to the neutrino mass diagram \Figref{fig:OneLoopNeutrinoMass} in the last section. 
LFV rare decays mediated by the flavor violating EW doublets $\varphi^{\prime (\prime)}$ are suppressed by one more loop order because of the necessity to couple to the neutrino sector VEVs. Hence, they only show up at two loop order, as shown in \Figref{fig:LFV-mu-to-e-gamma-2loop}. We will therefore not consider this diagram further.

\begin{wrapfigure}[9]{r}[0cm]{6cm}%[8]{h}[1cm]{6cm}%[tb]
\centering
\includegraphics[width=0.35\textwidth]{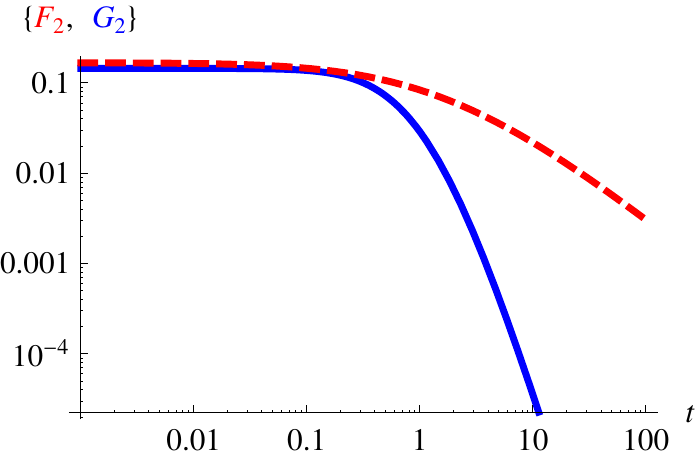}
\caption{The functions $F_2$(red) and $G_2$(blue).}\label{fig:FandG}
\end{wrapfigure}

Without any mass insertion along the $\eta$ line, a one-loop diagram of this type evaluates to~\cite{Ma:2001mr,Lavoura:2003xp,Kubo:2006kx}
\begin{align}
\mathrm{Br}(\mu\rightarrow e \gamma )= \frac{3 \alpha}{64 \pi(G_F m_0^2)^2}C^4,
\end{align}
where $m_0^2=\frac13 \left( M_1^2+M_2^2+M_3^2\right)$ and, using $x_J=(M^2_{\eta^+})_{JJ}/m_0^2$ and $h_{\alpha k J}=\frac{\partial \mathcal{L}_\nu}{\partial L_\alpha \partial S_k\partial {\hat\eta}_J}$, 
$$
C^2=\left\vert \sum_{i=1}^3 \sum_{J=1}^{9} h_{\mu i J}h^*_{e i J}x_J^{-2}F_2(M_S^2/(M^2_{\eta^+})_{JJ})  \right\vert 
\quad
\mathrm{and} 
\quad
F_2(t)=\frac{1-6t+3t^2+2t^3-6t^2\ln t}{6(1-t)^4}.\label{eq:F2def}
$$
In our model, we have $C^2=0$ for the symmetry reasons given above and there have to be mass insertions to generate flavor violating interactions. Note that this is a welcome feature since LFV processes 
of this type severely constrain models that generate neutrino masses radiatively~\cite{Kubo:2006kx}. 
This can be seen as the experimental constraint $\mathrm{Br}(\mu\rightarrow e \gamma)<2.4\cdot 10^{-12}$~\cite{Adam:2011ch} 
requires $C^4\sim 1.5 \cdot 10^{-8}$ for $M_{S}=m_0=100 \GeV$. The flavor symmetry automatically reduces $C^2$ by a factor $\left(\frac{\delta M_{\eta^+}^2}{M_{\eta^+}^2}\right)^2$.
In the limit $\left(\frac{\delta M_{\eta^+}^2}{M_{\eta^+}^2}\right)^2\ll 1$, the diagram \ref{fig:LFV-mu-to-e-gamma} can be computed explicitly and we find
\begin{align}
\mathrm{Br}(\mu\rightarrow e \gamma )= \frac{ \alpha}{16 \pi(G_F m_0^2)^2}\tilde{C}^4
\end{align}
where 
\begin{align}
\tilde{C}^2=\frac{1}{m_0^4}\left\vert \sum_{i=1}^3 \sum_{J,K,L=1}^{9} h_{\mu i J} \left(\delta M_{\eta^+}^2\right)_{JK} \left(\delta M_{\eta^+}^2\right)_{KL}  h^*_{e i L}F_4(M_S,M_J,M_K,M_L)  \right\vert 
\label{eq:expression-for-tilde-C}
\end{align}
and $F_4$ is a dimensionless loop integral, which we only give in the limit of degenerate $\eta$ masses
\begin{align}
G_2(t)&\equiv F_4(M_S=t m_0, M_J=m_0,M_K=m_0,M_L=m_0)\nonumber\\
&=\frac{1}{48(t^2-1)^{12}}\left[1-12 t^2-36 t^4+44 t^6+3 t^8-24(2 t^2+3)t^4 \ln t \right].
\label{eq:G2def}
\end{align}
The dimensionless functions $F_2$ and $G_2$ are plotted in \Figref{fig:FandG}. The explicit form of the sum in the expression \eqref{eq:expression-for-tilde-C} for $\tilde{C}^2$ is quite involved and will not be shown here, but it can be easily obtained using \Eqref{eq:deltaVeta} from the appendix. Here, we only comment on the generic size of the branching ratio. In general, the processes $\mu\rightarrow e \gamma$ and the radiative neutrino mass diagram break different approximate symmetries and it is therefore not necessarily the case that the smallness of neutrino masses implies a small branching ratio. This is also the case here. For example from Eq.~\eqref{eq:neutrino-mass-expressions}, one can read off that the smallness of neutrino mass could be due to very small values for $\lambda_1\approx \lambda_2\approx 10^{-9}$, with all other couplings being order one. Then the dominant contributions to $\tilde{C}^2$ would be of the type
\begin{align}
\tilde{C}^2\supset
\frac{G_2(t)}{m_0^4}\frac{1}{432} h_2\lambda_4 (bc-ad)\left[ - h_1\lambda_3 (ac+bd) + \omega^{2}\,h_2 \lambda_4 (bc-ad) \right]\;,
\end{align}
where we have again used the limit of degenerate masses $M_i=m_0$,
and could in principle be of order one. However, if we stick to the parts of parameter space where the smallness of neutrino mass is due to many moderately small couplings $h_i,\lambda_i\sim\Order{0.01-0.1}$ and $m_{0}\sim\Order{\TeV}$, $\ev{\chi},\ev{\phi_i}\sim\Order{100\GeV}$ (as discussed below \eqref{eq:neutrino-mass-expressions}) instead of one very small coupling, the branching ratio is heavily suppressed by ${\tilde C}^4\sim (10^{-9}-10^{-13})^2$. These natural parameter values thus give an appealing explanation of both the smallness of neutrino masses and the suppression of LFV decays.

\addtocontents{toc}{\protect\setcounter{tocdepth}{1}}
\subsection{LFV Decays $l_i\rightarrow lll$}
\addtocontents{toc}{\protect\setcounter{tocdepth}{2}}
\label{sec:LFVintoLeptons}
Another class of processes that are of interest for our model are rare flavor violating decays of the type $\mu\rightarrow eee$. 
As in the case of the processes $f_i\rightarrow f_j \gamma$ the allowed decay channels are restricted by the flavor symmetry. If we do 
not consider the heavily suppressed diagrams that couple to VEVs in the neutrino sector, it is clear that the process $\mu\rightarrow eee$
is not allowed by the $Z_3$ symmetry of the charged lepton sector and the most constraining process is given by $\tau^-\rightarrow \mu^- \mu^- e^+$ . 

\begin{figure}[tb]\centering
\begin{subfigure}{.35\textwidth}
\includegraphics[width=\textwidth]{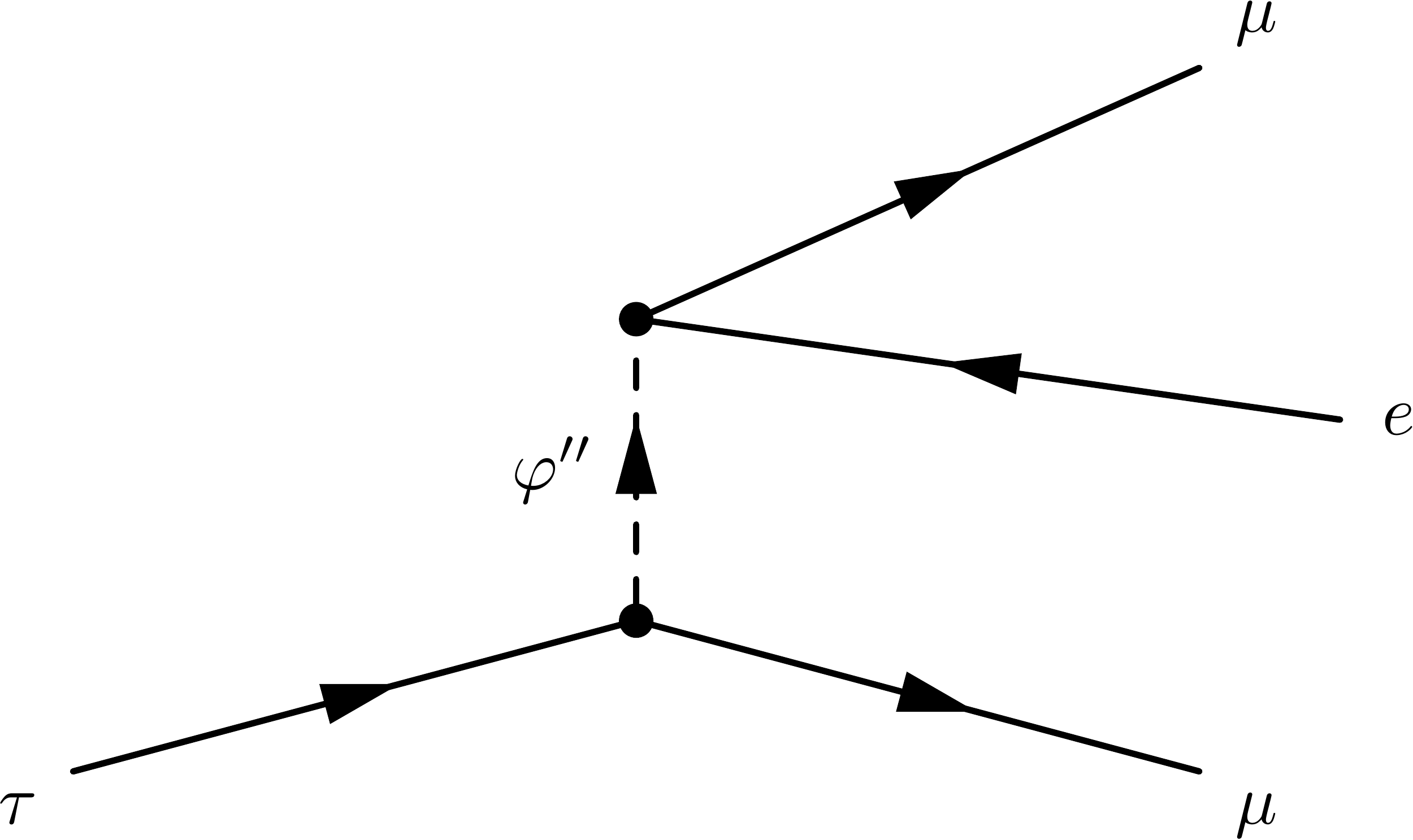}
\caption{Tree level contribution of $\varphi^{\prime\prime}$\label{fig:Tau2MuMuEvarphi}}
\end{subfigure}
\hspace{2 cm}
\begin{subfigure}{.35\textwidth}
\includegraphics[width=\textwidth]{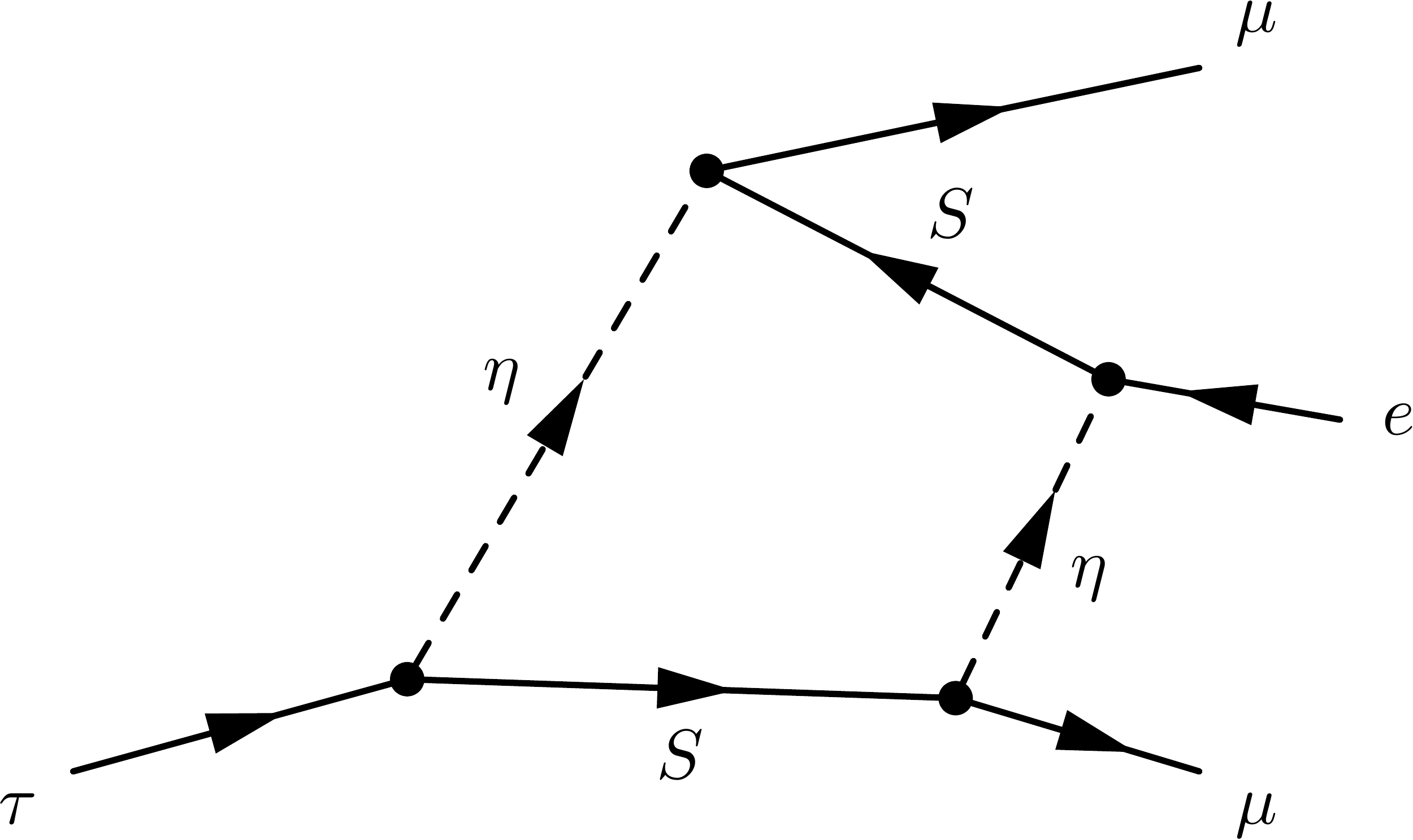}
\caption{One loop contribution of $S$-$\eta$ \label{fig:Tau2MuMuEeta}}
\end{subfigure}
\caption{Lepton Flavor Violating rare decay $\tau^-\to\mu^-\mu^- e^+$. }\label{fig:LFV-decays}
\end{figure}

This process can be mediated at tree-level by the neutral components of $\varphi^{\prime \prime}$ as depicted in \Figref{fig:Tau2MuMuEvarphi} and its branching ratio is given by~\cite{Ma:2010gs,Adelhart-Toorop:2010uq}
\begin{align}
 \mathrm{Br}(\tau^-\rightarrow \mu^- \mu^- e^+)=\left(\frac{36 m_\tau^2 m_\mu^2}{ M_0^4} \right) \mathrm{Br}(\tau\rightarrow \mu \nu \nu)& 
 =1.7\cdot10^{-8}  \left(\frac{62\GeV}{M_0}\right)^4 
\end{align}
where we have used $\mathrm{Br}(\tau\rightarrow \mu \nu \nu)=0.174$. Compared to the experimental upper bound of $1.7\cdot 10^{-8}$~\cite{PDG:2012}, the effective mass\footnote{In~\cite{Ma:2010gs} $\lambda_{\chi A}=0$ was assumed, which implies $\alpha=\pi/4$.} 
\begin{align}
\frac{1}{M_0^4}=\left[ \frac{\sin^2\alpha}{m_{{\Phi_1}}^2}+\frac{\cos^2\alpha}{m_{{\Phi_2}}^2}\right]^2
\label{eq:eff-mass}
\end{align}
is thus only weakly constrained. All other processes mediated by $\varphi^{\prime(\prime)}$ are further suppressed by $y_e y_\tau$ or $y_\mu y_e$.
Rare LFV processes mediated by these fields are therefore naturally suppressed by smallish Yukawa couplings and do not put a serious constraint on the model. 
\\
Let us also estimate the magnitude of the diagram in \Figref{fig:Tau2MuMuEeta} mediating $\tau\rightarrow \mu \mu e$, as this diagram may in principle be larger because it is not suppressed by Yukawa couplings that are known to be small. 

To get an estimate, we work in the limit of degenerate $\eta$ masses $M_1=M_2=M_3=m_0$ and find
$$
\Gamma(\tau^-\rightarrow \mu^- \mu^- e^+ )\sim \left\vert\frac{1}{16 \pi^{2}}\sum_{j,k=1}^{9}\sum_{i,l=1}^{3} h_{\tau ij}h_{\mu ik}^* h_{e lk}h_{\mu lj}^*\frac{H(M_S/m_0)}{ m_0^2}\right\vert^2,
$$
where $H(M_S/m_0)$ is a dimensionless loop integral and 
$$
\mathrm{Br}(\tau^-\rightarrow \mu^- \mu^- e^+)=\mathrm{Br}(\tau\rightarrow \mu \nu \nu)\frac{\Gamma(\tau^-\rightarrow \mu^- \mu^- e^+ )}{\Gamma(\tau^-\rightarrow \mu^- \overline{\nu}_\mu \nu_\tau)}\;.
$$
Evaluating the sum, we find $\sum_{j,k=1}^{9}\sum_{i,l=1}^{3} h_{\tau ij}h_{\mu ik}^* h_{e lk}h_{\mu lj}^*=\frac{1}{27}\left( h_1^4-h_1^2h_2^2+h_2^4\right)$ and the experimental bound
\begin{align}
\mathrm{Br}(\tau^-\rightarrow \mu^- \mu^- e^+ )= \left\vert \left( \frac{159\GeV}{m_0}\right)^2(h_1^4+h_2^4-h_1^2h_2^2)  H(M_S/m_0)\right\vert^2 \cdot 1.7\cdot 10^{-8}
\end{align}
can easily be evaded even for small values of $m_0\sim 178 \GeV\approx 308/\sqrt{3} \GeV$ (which would give the correct dark matter relic abundance of $\eta$ in the degenerate limit we are considering here, as will be discussed in \Secref{sec:DM-pheno}) and order one Yukawas (assuming $H(M_S/m_0)\sim 1$). For the parameter ranges preferred by one-loop neutrino mass generation, i.e. $h_i\sim 0.1$, the expected branching ratio is too small to expect a signal in next-generation experiments. In summary, we can conclude that the flavor symmetry effectively protects against lepton flavor violating interactions.

\addtocontents{toc}{\protect\setcounter{tocdepth}{1}}
\subsection{Anomalous Magnetic Moment of Muon}
\addtocontents{toc}{\protect\setcounter{tocdepth}{2}}
\label{sec:anomMagMom}
Let us now briefly discuss the anomalous magnetic moment of the muon. The contribution from
the exchange of the neutral component of $\varphi^{\prime \prime}$ should give the largest contributions, as it is proportional to the tau Yukawa coupling squared. 
It has been calculated previously~\cite{Ma:2001fk} and amounts to
\begin{align}
\Delta a_\mu= \frac{G_F m_\tau^2}{2\sqrt{2}\pi^2}\left( \frac{m_\mu^2}{M_0^2}\right)=1.5 \cdot 10^{-12} \left(\frac{100 \GeV}{M_0}\right)^2,
\end{align}
which is negligible and cannot account for the reported deviation of $(290\pm90)\times 10^{-11}$~\cite{Bennett:2006yq,Jegerlehner:2009lr}, from the Standard Model. The charged components of $\eta$ 
also contribute to the anomalous magnetic moment of the muon, with a strength given by~\cite{Jegerlehner:2009lr,Ma:2001mr}
\begin{align}
\Delta a_\mu&=-\frac{m_\mu^2 }{3(4\pi^2)^2}\left[\frac{h_1^2}{M_1^2} F_2\left(\frac{M_S}{M_{1}}\right)+\frac{h_2^2}{M_2^2} F_2\left(\frac{M_S}{M_{2}}\right) \right]\nonumber\\
&=-1.8\times 10^{-12}\sum_i \left(\frac{h_i}{0.1}\right)^2\left(\frac{100\GeV}{M_i}\right)^2 \left(\frac{ F_2(\frac{M_S}{M_i})}{F_2(1)}\right).
\end{align}
This therefore gives a very mild constraint on the masses and Yukawa couplings of the $\eta$'s. In the preferred parameter space for neutrino mass generation, this contribution is negligible. Note that the contribution goes in the opposite direction
of the reported excess and it can therefore not be used to explain it~\cite{Ma:2001mr}.

%%%%%%%%%%%%%%%%%%%%%%%%%%%%%%%%%%%%%%%%%%%%%%%%%%%%%%%%%%%%%%%%%%%55
%%%%%%%%%%%%%%%%%%%%%%%%%%%%%%%%%%%%%%%%%%%%%%%%%%%%%%%%%%%%%%%%%%%55
%%%%%%%%%%%%%%%%%%%%%%%%%%%%%%%%%%%%%%%%%%%%%%%%%%%%%%%%%%%%%%%%%%%55

%\input{dark-matter-v1}
\section{Dark Matter}
\label{sec:DM}
In this section we discuss dark matter candidates of the model and their phenomenology.
\subsection{Dark Matter Candidates and their Stability}
\label{sec:DM-stab}
To start off the discussion of possible dark matter candidates in our model, let us dwell on the remnant symmetries left over after symmetry breakdown. While the $Q_8\rtimes A_4$ part of the symmetry group is completely broken\footnote{There have been several studies of dark matter, which is stabilized by a remnant subgroup of a flavor symmetry~\cite{Hirsch:2010ru,*Meloni:2010sk,*Toorop:2011ad,*Meloni:2011cc,*Boucenna:2011tj,*Boucenna:2012qb,*Lavoura:2012cv}, while it is completely broken in our model.}, there is a $Z_2$ symmetry given by 
\begin{align}
\mathcal{R}:\qquad L&\rightarrow -L& \ell^c&\rightarrow -\ell^c & \eta_i\rightarrow -\eta_i\; ,
\end{align}
which is the $(-1)^{L^\prime}$ remnant of the auxiliary $Z_4$ symmetry $i^{L^\prime}$, where $L^\prime=L+N_\eta$ is the generalized lepton number symmetry that is the sum of the usual SM lepton number with the $\eta$ number $N_\eta$. At the renormalizable level after symmetry breaking, there is another $Z_2$ symmetry of the model given by 
\begin{align}
\mathcal{A}:\qquad S &\rightarrow -S & \eta_i\rightarrow -\eta_i\;.
\end{align}
This is purely an accidental symmetry that emerges due to the particle content and the requirement of renormalizability and not a remnant of some symmetry we have imposed on the model. The reason why it emerges can be traced back to the fact that the SM fermions as well as $\chi$ transform only under the generators $S$ and $T$ that form the subgroup $A_4$ and thus there are no operators of the form $\varphi \mathcal{O}_{A_4}$, where $\varphi$ is a field transforming non-trivially under $X$ (e.g. fields transforming as $\MoreRep{3}{i}$ with $i\neq1$ such as $S$ and $\eta$) and $\mathcal{O}_{A_4}$ is an arbitrary operator formed by fields transforming under $A_4$.
%These two symmetries in tandem make dark matter stable. Note that the remnant symmetry $\mathcal{R}$ alone would not be sufficient, as e.g. the decay of the lightest particle contained in $\eta_i$ to neutrinos and the neutral CP-even component of the Higgs would be possible.
The symmetry $\mathcal{A}$ makes the lightest component of $S$ and $\eta$ stable, which implies that the dark matter candidate is either fermionic or bosonic. This symmetry, however, is only an accidental symmetry and there is thus no reason for higher dimensional operators to respect this symmetry. Such a higher dimensional operator $\mathcal{O}$ with $\mathcal{A}[\mathcal{O}]\neq \mathcal{O}$ would lead to a decay of the dark matter candidate. On the contrary, all higher dimensional operators have to respect the symmetry $\mathcal{R}[\mathcal{O}]= \mathcal{O}$, as this symmetry is a remnant of an exact symmetry and is therefore also exact. We will now show that this requirement pushes up the dimensionality of the higher dimensional decay operators to a level where the dark matter candidate is stable for all practical purposes. Since the discussion depends on whether the dark matter candidate stems from $\eta$ or from $S$, we discuss the two possibilities in turn.

\subsubsection{Scalar DM}
Any effective operator that would mediate a decay of the lightest component of $\eta_i$ has to be of the form 
\begin{align}\label{eq:DMdecayOperator}
\mathcal{O}=\eta_i \mathcal{O}_{SM}^{\Delta L=1} \vev{\mathcal{O}_{\phi_k \phi_l}}
\end{align}
where $\vev{\mathcal{O}_{\phi_k \phi_l}}$ is built out of SM-singlet flavon fields and transforms even under $\mathcal{R}$. As $\eta$ is odd under $\mathcal{R}$, the operator $\mathcal{O}_{SM}^{\Delta L=1} $, which is built up of SM fields, has to be also odd under $\mathcal{R}$ to make the complete operator invariant. Obviously the complete operator $\mathcal{O}$ is odd under the accidental symmetry $\mathcal{A}$ and thus mediates DM decay.

Since $\mathcal{R}$ acts upon SM particles as the discrete subgroup of lepton number $(-1)^L$, the operator $\mathcal{O}_{SM}^{\Delta L=1}$ has to violate lepton number by an odd unit and has to transform as an electroweak doublet. The lowest dimensional operators in the SM arise at dimension six and violate $L$ by one unit
(See~\cite{Grzadkowski:2010ve} for a recent review of gauge invariant dimension 6 operators.)
\begin{subequations}
\label{eq:DMdecaySMOps}
\begin{align}
&\label{eq:dim6Ops}
L u^c d^c d^c
\quad\quad\bar L \bar d^c \bar d^c \bar d^c
\quad\quad L\bar Q \bar Q d^c
\quad\quad \bar e^c \bar Q d^c d^c
\\
&\label{eq:dim7Ops}
\chi^\dagger LQQQ
\quad\quad
\chi^\dagger e^c u^c u^c d^c
\quad\quad
\chi^\dagger \bar L \bar Q  u^c d^c 
\quad\quad
\chi^\dagger \bar e^c Q Q \bar u^c 
\quad\quad
\chi^\dagger L Q \bar u^c\bar d^c\;.
\end{align}
\end{subequations}
All dimension 6 operators in \Eqref{eq:dim6Ops} break baryon number by one unit, $B-L$ by two units and preserve $B+L$.
The dimension 7 operators in \Eqref{eq:dim7Ops} on the other hand break baryon number by one unit, preserve $B-L$ and break $B+L$ by two units. They are formed by adjoining $\chi$ to a dimension 6 proton decay operator. Since baryon number is an accidental symmetry in our model (in the same way as in the Standard Model), these operators are never generated\footnote{Except through instantons and sphalerons, which do not play a role here, in the same way as in the SM.} within the model and thus dark matter is stable within the model. They rather parameterize some baryon number violating physics, which from proton decay experiments is pushed to scales of the order of $\Lambda_B \approx 10^{16} \GeV$.

To form a singlet under the flavor symmetry, the second operator $\mathcal{O}_{\phi_k\phi_l}$ is needed to make the total operator $\mathcal{O}$ a singlet under the flavor symmetry, as $\eta_i$ transforms under $X$ while $\mathcal{O}_{SM}^{\Delta L=1} $ does not. It has to be composed of an even number of flavons $\phi_k$, as under the $Z_2$ subgroup generated by \footnote{This element generates the center of the group and thus commutes with all group elements.} $X^2$ only $\phi_k$ transforms non-trivially.

If we assume the presence of baryon number violating operators at scale $\Lambda_B$, the dark matter candidate $\eta$ decays into quarks and one lepton. Under the assumption that the flavor part of the operator is related to the breaking of the flavor symmetry $\Lambda_F$, a DM decay operator formed by a dimension 6 SM operator $\mathcal{O}_{SM}^{\Delta L=1}$ is suppressed by $\Lambda_B^3$:
\begin{equation}
\frac{\eta_i \mathcal{O}_{SM}^{\Delta L=1}}{\Lambda_B^3} \frac{\vev{\phi_k \phi_l}}{\Lambda_F^2}\;.
\end{equation}
Hence, the lifetime of DM can be estimated to be
\begin{equation}
\Gamma^{-1}\sim \frac{8\pi \Lambda_B^6}{m_\eta^7}\left(\frac{\Lambda_F^2}{\vev{\phi_k\phi_l}}\right)^2=1.9\cdot 10^{45} \mathrm{Gyr} \left(\frac{\Lambda_B}{10^{16} \GeV}\right)^6 \left(\frac{100 \GeV}{m_\eta}\right)^7 \left(\frac{\Lambda_F^2}{\vev{\phi_k\phi_l}}\right)^2\;
\end{equation}
and the dark matter candidate is thus stable even on cosmological time-scales, if one assumes `traditional' values for the scale of baryon number violating physics. However the operators in \Eqref{eq:dim6Ops} are not those directly tested in proton decay experiments and the physics of baryon number violation might be such that the operators in \Eqref{eq:dim6Ops} are suppressed by a smaller energy scale than the one responsible for baryon decay. We will come back to the issue of induced proton decay at the end of the subsection, but now we want to turn the logic around and derive bounds on $\Lambda_B$ and $\Lambda_F$ from the fact that dark matter is still around.

Decaying DM models are constrained by WMAP to $\Gamma^{-1}\geq 123\, \mathrm{Gyr}$ at 68\% C.L.~\cite{Ichiki:2004vi} and WMAP+SN Ia to $\Gamma^{-1}\geq 700\, \mathrm{Gyr}$ at 95.5\% C.L.~\cite{Gong:2008gi}. Furthermore, decaying DM is constrained by possible neutrino final states~\cite{PalomaresRuiz:2007ry}, which serve as a conservative limit, since neutrinos are the least detectable SM particles. The exact bound depends on the DM mass ranging from $10^{22}s=10^8 \mathrm{Gyr}$ at $\mathcal{O}(1\GeV)$ and increasing almost linearly on a log-log plot to $10^{28}s\approx 10^{14} \mathrm{Gyr}$ at $\mathcal{O}(100\TeV)$. Diffuse gamma ray constraints from Fermi data yield a limit of $\Gamma^{-1}\gtrsim 10^{26}s\approx 10^{12} \mathrm{Gyr} $~\cite{Cirelli:2009dv} for the decay into a pair of charged leptons. Here, DM decays into one lepton and quarks, which might lead to further softer leptons in the final state. Hence the bounds do not directly apply, but we will use it to obtain an order of magnitude estimate for the suppression scale of the lowest order DM decay operator in \Eqref{eq:DMdecayOperator}. Using the limit from diffuse gamma rays with $\Gamma^{-1}\gtrsim10^{26} s$ as a benchmark value, we obtain a limit on the suppression scale of
\begin{equation}
\left(\Lambda_B^3\Lambda_F^2\right)^{1/5}\gtrsim 6 \cdot 10^{7}\, \mathrm{GeV} \left(\frac{m_\eta}{1\TeV}\right)^{7/10}\left(\frac{\vev{\phi_k\phi_l}}{(100\GeV)^2}\right)^{1/5}\;.
\end{equation}
Because of the high dimensionality of the operator, the bound on the suppression scale $\Lambda_{B,F}$ does not depend strongly on the bound on the lifetime. 

All of the operators in \Eqref{eq:DMdecaySMOps} lead to DM induced proton decay~\footnote{Induced proton decay has been studied in the context of asymmetric DM~\cite{Davoudiasl:2011fj}. However, their analysis does not apply in our case, because the induced proton decay is mediated via a different operator with different kinematics, since one of the final state particles has a non-negligible mass of the order of the proton mass.} into a final state lepton and final state mesons
\begin{equation}
\eta_i + N \rightarrow L+ M\;.
\end{equation} 
As the proton as well as the DM are non-relativistic and they annihilate at rest, the induced proton decay leads to similar kinematics as in the ordinary proton decay, but the total rest energy $E\sim m_{\eta} + m_N\approx m_{\eta}$ is much larger compared to the ordinary proton decay with $E\sim m_N$. Hence, the final state particles appear to originate from the decay of a much heavier particle and the experimental signatures change. Therefore, the existing limits on proton decay are not directly applicable. However, in generic GUT models, for example, the operators given in Eqs. \eqref{eq:dim6Ops}, \eqref{eq:dim7Ops} and the proton decay operators are generated at the same energy scale.

\subsubsection{Fermionic DM}
Similarly to scalar DM consisting of the lightest component of $\eta_i$, $S$ can decay via higher-dimensional operators. They are generally of the form
\begin{equation}
S \mathcal{O}_{SM} \braket{\mathcal{O}_{\phi_k\phi_l}}\;,
\end{equation}
where $\mathcal{O}_{SM}$ transforms like a spin $\frac12$ fermion, which is a singlet under the SM group, but transforms non-trivially under the flavor symmetry\footnote{Note that S transforms under the symmetry generator $X$, while $\mathcal{O}_{SM}$ does not. Therefore the operator $\braket{\mathcal{O}_{\phi_k\phi_l}}$ is needed to form a singlet.}. The lowest dimensional operators $\mathcal{O}_{SM}$ emerge at dimension $\frac92$

\begin{align}
\label{eq:dim6OpsS}
u^c d^c d^c
\qquad \bar Q \bar Q d^c\qquad
\chi Q\bar u^c \bar d^c
\qquad\quad\chi QQQ.
\end{align}
Note that these operators transform trivially under $\mathcal{R}$, as does $S$. All of these operators violate baryon number by one unit and therefore, they lead to induced proton decay. However, the kinematics is quite different compared to ordinary proton decay, because the lowest order operators do not contain a final state lepton. 

Similarly to the scalar case, there are bounds from astrophysical observations. As DM decay only arises at dimension 8, the bound on the suppression scale does not depend strongly on the exact bound on the lifetime. Therefore, we again make a rough estimate of the bound on the suppression scale by using the same lifetime as in the scalar case and we obtain
\begin{equation}
\left(\Lambda_B^2\Lambda_F^2\right)^{1/4}\gtrsim 9 \cdot 10^{8}\, \mathrm{GeV} \left(\frac{m_\eta}{1\TeV}\right)^{5/8}\left(\frac{\vev{\phi_k\phi_l}}{(100\GeV)^2}\right)^{1/4}\;
\end{equation}
due to the lower dimensionality of the DM decay operator.

\subsection{Dark Matter Phenomenology}
\label{sec:DM-pheno}
We now give a brief overview of the phenomenology of the two different dark matter candidates. We will estimate the DM abundance and detection possibilities for the different scenarios and show that there is a region of parameter space where the correct abundance can be obtained. A detailed calculation is beyond the scope of the present work. Again, we discuss the different dark matter candidates separately. 

\subsubsection{Scalar DM}

The scalar dark matter candidate is a component of an inert EW doublet. Therefore, we are going to translate the analysis for scalar multiplet DM done in~\cite{Hambye:2009pw} to our setup. A detailed analysis would require the precise calculation of the $\eta_i$ mass matrices. We assume that one of the triplets $\eta_i$ is sufficiently lighter than the other two, such that we do not have to take them into account during freeze-out of DM, i.e. they have to be at least $20\%$ heavier than the DM candidate~\cite{Griest:1990kh}. In the following, we will denote the triplet containing the DM candidate by $\eta_{DM}$ with direct mass term $M_{\eta_{DM}}$.
We are going to assume, as we did previously in the section about the neutrino masses, that the direct mass term $M_{\eta_{DM}}$ dominates over all mass terms induced by VEVs. Hence, the DM mass is approximately given by the direct mass term $M_{\eta_{DM}}$. 
In the limit that the mass splittings are below 1\%, we can neglect the annihilations via other scalars and concentrate on the pure gauge (co)annihilation channels. Following~\cite{Hambye:2009pw}, there is an upper bound on the DM mass of an inert doublet of $m^*=534\pm25\GeV (3\sigma)$ from overclosing the universe in this limit. The correct DM abundance is obtained for $m^*$. As $\eta_{DM}$ is in a triplet representation of $Q_8\rtimes A_4$, there are three almost degenerate doublets, which all contribute to the DM density equally. Therefore, the upper bound on the DM mass is lowered by approximately a factor of $\sqrt{3}$ to $m^*_\eta\approx 308\GeV$, which is consistent with direct searches for scalar particles, as discussed in \Secref{sec:collider}.

Today, the mass splitting between DM and the next-to lightest particles forbids gauge interactions kinematically due to the small DM velocities, unless it is tuned to be very small ($\lesssim \Order{100}\keV$), and DM can only be detected via the couplings to scalars, specifically via the Higgs portal. The spin-independent cross section for scattering of DM off the neutron is given by~\cite{Andreas:2008xy}
\begin{equation}
\sigma_n\approx\frac{|\lambda_L|^2}{\pi} \frac{\mu^2}{M_{DM}^2}\frac{m_p^2}{m_H^4} f^2\approx 2.7\cdot 10^{-48} \left(\frac{\lambda_L}{0.01}\right)^2\left(\frac{300\GeV}{M_{\eta_{DM}}}\right)^2\left(\frac{125\GeV}{m_H}\right)^4 \left(\frac{f}{0.3}\right)^2\mathrm{cm}^2
\end{equation}
with $\lambda_L$ being the coupling of DM to the Higgs, $\mu$ the reduced mass of the DM-neutron system, $m_p$ the mass of the nucleon, $m_H$ the mass of the Higgs and $f$ parametrises the nuclear matrix element, $0.14<f<0.66$, which we took from~\cite{Andreas:2008xy}. The estimated cross section is well below the current experimental limits by XENON100~\cite{Aprile:2011hi}, which is the most sensitive DM direct detection experiment in this mass region.

Note, the discussed parameter point is only an example that proves the possibility of obtaining the correct DM relic density. For larger mass splittings, the annihilation via scalar interactions cannot be neglected in the calculation of the DM relic abundance and the direct detection cross section is enhanced.

\subsubsection{Fermionic DM}
For the discussion of the fermionic DM candidate contained in $S$, we follow the discussion in~\cite{Kubo:2006kx} to show that it is possible to obtain the correct relic abundance. For completeness, we repeat the relevant steps with the necessary changes. At tree-level, there is only the mass term $\sqrt{3}M_S S S=M_S(S_1^2+S_2^2+S_3^2)$ and thus all components of $S$ are degenerate. At loop-level this degeneracy is lifted and for concreteness we here take $M_{\tilde S_3}\gtrsim M_{\tilde S_2}\gtrsim M_{\tilde S_1}$, where $\tilde S_i$ are mass eigenstates. The states $\tilde S_{2,3}$ can decay into $\tilde S_1$ and leptons by the interchange of $\eta$ and thus at the present time only $\tilde S_1$ is around. However, due to the near degeneracy, the freeze-out of all three species runs in parallel. Coannihilation processes of the type $S_i S_j\rightarrow \mathrm{SM}$ with $i\neq j$ are suppressed in comparison to annihiliation processes $S_i S_i\rightarrow \mathrm{SM}$, because they require an additional mass insertion along the $\eta$ line. It is thus a very good approximation to consider the freeze-out of each component separately and the total relic abundance is thus just given by the sum of the abundances of $S_1$, $S_2$ and $S_3$.

\begin{figure}
\begin{center}
\includegraphics[width=.7\textwidth]{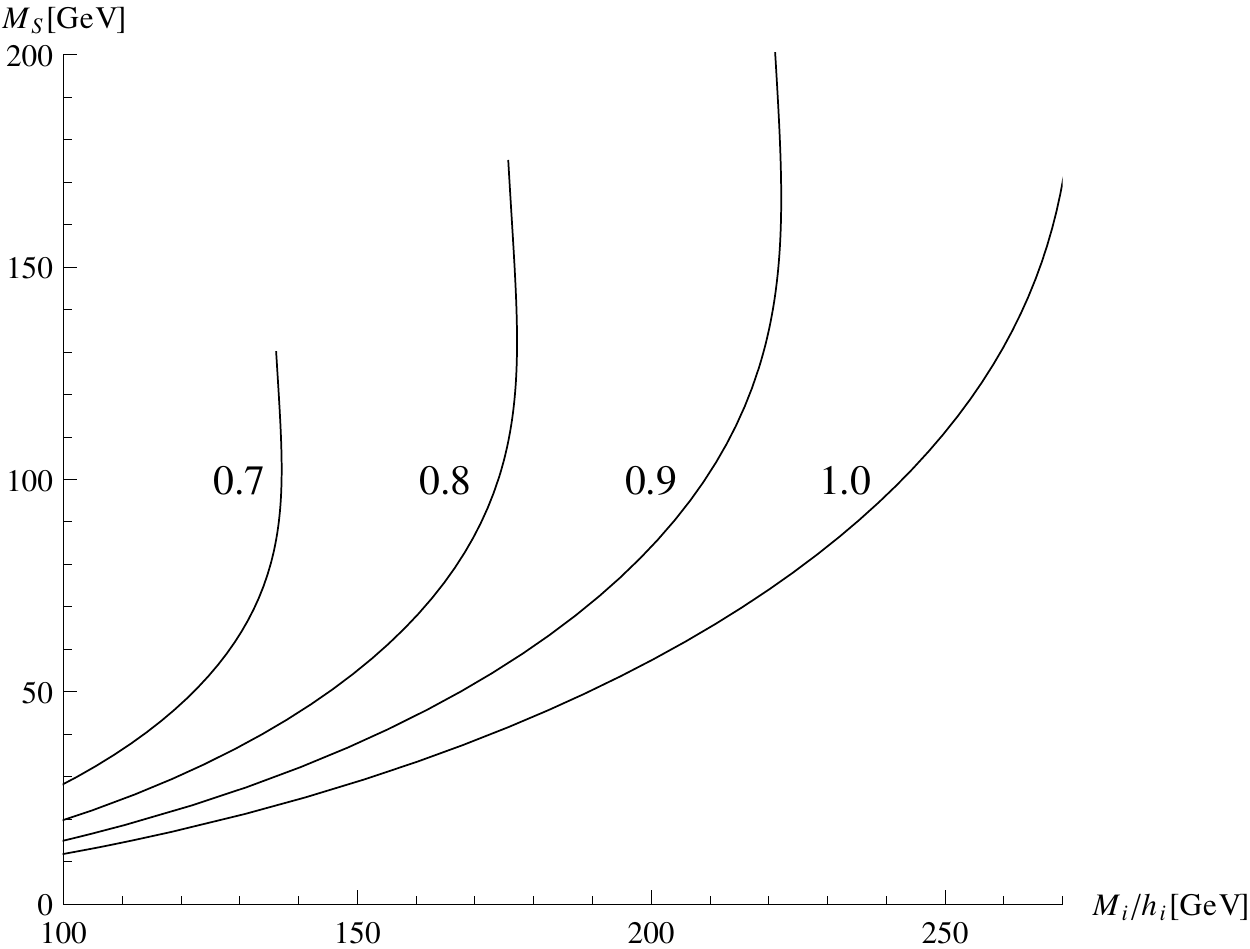}
\caption{Contour lines for different values of $h_1=h_2$ with the correct DM abundance $\Omega_S h^2=0.12$.\label{fig:fermionicDM}}
\end{center}
\end{figure}

The annihilation cross section for each $S_k$ into leptons in the limit of vanishing lepton masses and scalar mass splittings~\cite{Griest:1988ma} is given by
\begin{align}
\ev{\sigma v}&=b v^2+\Order{v^4},&
b&=\sum_{i=1,2}\frac{h_i^4 r_i^2 \left(1-2r_i+2r_i^2\right)}{24\pi M_S^2},&
r_i&=\frac{M_S^2}{M_i^2+M_S^2}\;.
\end{align}
In the limit of $M_S\ll M_i$, the expression for the p-wave simplifies to
\begin{equation}
b=\frac{M_S^2}{24\pi} \sum_{i=1,2}\left(\frac{h_i}{m_i}\right)^4 \;,
\end{equation}
i.e. the cross section scales with $(h_i/m_i)^{4}$.
The relic density of the SM singlets $S$, taking into account the mass degeneracy of the components of $S$, can then be obtained from~\cite{Griest:1989zh}
\begin{align}\label{eq:omegaS}
\Omega_S h^2=\frac{n_{S}^0 M_S}{\rho_c} h^2,
\end{align}
with $n_{S}^0$ being the number density of $S$ today, which is
\begin{equation}\label{eq:density}
(n_S^{0})^{-1}=(\sum_k n_{S_k}^{0})^{-1}=(3 n_{S_k}^{0})^{-1}=\frac{0.088\, g_*^{1/2}M_{Pl} M_S 3b}{x_f^2 s_0}\;,
\end{equation}
where $s_0=2970/\mathrm{cm}^3$ is today's entropy density, the critical density is $\rho_c=3H^2/(8\pi G)=1.05\cdot 10^{-5} h^2 \mathrm{GeV}/\mathrm{cm}^3$, the Planck mass $M_{Pl}=1.22\cdot 10^{19}\GeV$ and the dimensionless Hubble parameter $h$.
At the freeze-out temperature, the ratio $x_f=M_S/T$ is determined by
\begin{equation}
x_f=\ln \frac{0.0764 M_{Pl} (6 b/x_f) c (2+c) M_S}{(g_* x_f)^{1/2}}
\end{equation}
with the effective number of degrees of freedom $g_*$ at freeze-out. After eliminating the cross section with \Eqref{eq:density} and \Eqref{eq:omegaS}, we obtain
\begin{equation}\label{eq:xf}
x_f=\ln \frac{1.74 x_f^{1/2} s_0 h^2c (2+c) M_S}{g_* (\Omega_S h^2) \rho_c}.
\end{equation}
Following the discussion in~\cite{Griest:1989zh,Kubo:2006kx}, we rewrite \Eqref{eq:omegaS} and \Eqref{eq:xf} as
\begin{subequations}
\begin{align}
\left[\frac{M_S}{\GeV}\right]&=1.95\cdot 10^{-8} x_f^{-1/2} e^{x_f}\left[\frac{\Omega_d h^2}{0.12}\right]\label{eq:Mnum},\\
\left[\frac{b}{\GeV^{-2}}\right]&=7.32\cdot 10^{-11} x_f^2\left[\frac{\Omega_d h^2}{0.12}\right],
\end{align}
\end{subequations}
using $g_*^{1/2}=10$ and $c=1/2$. We solve these equations numerically for fixed values of $h_1=h_2$ and $M_1=M_2$ and show the resulting contour lines with the correct DM relic abundance in the plane $M_1/h_1=M_2/h_2$ vs.~$M_S$ in \Figref{fig:fermionicDM}.
Hence, it is possible to obtain the correct DM relic abundance for fermionic DM, although large Yukawa couplings $h_i$ are required. Similarly to the scalar DM scenario, we expect the cross section to raise with non-vanishing mass splittings of the scalars $\eta_i$, which allows for smaller Yukawa couplings $h_i$.

%%%%%%%%%%%%%%%%%%%%%%%%%%%%%%%%%%%%%%%%%%%%%%%%%%%%%%%%%%%%%%%%%%%55
%%%%%%%%%%%%%%%%%%%%%%%%%%%%%%%%%%%%%%%%%%%%%%%%%%%%%%%%%%%%%%%%%%%55
%%%%%%%%%%%%%%%%%%%%%%%%%%%%%%%%%%%%%%%%%%%%%%%%%%%%%%%%%%%%%%%%%%%55
%\input{quarks-v1}
\section{Extension to Quark Sector}
\label{sec:quarks}
So far we restricted ourselves to the discussion of the flavor structure in the lepton sector. Given the different structures in the lepton and quark sector, one might wonder whether and how this model can be extended to the quark sector.
In the following, we will discuss a few simple possibilities to incorporate the quark sector without enlarging the flavor group. It is necessary to specify how quarks transform under the flavor symmetry as this will to a certain extent determine the collider signatures of the model. 
Alternatively, it is interesting to look for a group extension of the flavor group, which preserves the structure in the lepton sector, but allows for new structure in the quark sector~\cite{Holthausen:2011vd}. Here, a viable extension of the full flavor group $Q_8\rtimes A_4$ is the group $Q_8\rtimes T^\prime\cong\SG{192}{1022}$~\cite{Holthausen:2011vd} being the analogue of the extension of $A_4$ to $T^\prime$, which has been used to explain the flavor structure of quarks and leptons simultaneously~\cite{Frampton:1994rk,*Aranda:2000tm,*Feruglio:2007uu,*Frampton:2007et,*Frampton:2008bz,*Eby:2008uc,*Eby:2011ph}. 
A detailed discussion of quark flavor observables is postponed to future work.

\subsection*{Quark Sector Mirroring the Lepton Sector}
\label{sec:quarks-simplest}
We can use the same assignment for the quarks as for the leptons with respect to $(Q_8\rtimes A_4)\times Z_4$, i.e. 
\begin{align}
Q&\sim (\MoreRep{3}{1},1)\;,&% \\\nonumber
u^c+c^c+t^c&\sim (\MoreRep{1}{1}+\MoreRep{1}{2}+\MoreRep{1}{3},1)\;,&
d^c+s^c+b^c&\sim (\MoreRep{1}{1}+\MoreRep{1}{2}+\MoreRep{1}{3},1)\;.
\end{align}
This assignment leads to the following Yukawa couplings in the Lagrangian
\begin{equation}
-\mathcal{L}_q = y_u\, Q \chi u^c 
+y_c\, Q \chi c^c  
+y_t\, Q \chi t^c 
+y_d\, Q \tilde{\chi} d^c 
+y_s\, Q \tilde{\chi} s^c  
+y_b\, Q \tilde{\chi} b^c 
  +\hc\;,
\end{equation}
which amount to the mass matrices of the quarks
\begin{align}
M_U&=\frac{v}{\sqrt{2}} \Omega_T^*\diag(y_u,y_c,y_t)& &\mathrm{and}&
M_D&=\frac{v}{\sqrt{2}} \Omega_T^*\diag(y_d,y_s,y_b)\;.
\end{align}
Hence there is no mixing in the quark sector, i.e. the CKM mixing matrix $V_{CKM}=V_d^\dagger V_u =\mathbbm{1}$, which is a good leading order approximation to the CKM mixing.
The Cabibbo angle can be produced by a cross-talk of operators from the neutrino sector~\cite{He:2006dk}, e.g. the operator $(Q \tilde{\chi} d^c)_{\MoreRep{1}{2}} (\phi_1 \phi_2)^2/M^4$ leads to a non-vanishing Cabibbo mixing angle. It has to be of the order $(\phi_1 \phi_2)^2/(M^4)\sim 10^{-4} (m_s/95\MeV)$, in order to generate a large enough mixing in the down-type quark sector to explain the Cabibbo angle. Within the model, the operator can be generated at one loop with $\varphi^{\prime(\prime)}$ running in the loop. However, the contribution turns out to be too small and a different mechanism is required to generate this operator.

Flavor changing neutral currents are naturally suppressed at the leading order, since there is a selection rule $\Delta D \Delta S \Delta B=\pm 2$ as well as $\Delta U \Delta C \Delta T=\pm2$ 
in the flavor basis for four Fermi operators similarly to the lepton sector.
It has been claimed in~\cite{Ma:2010gs} that leptonic Kaon decays result in a relatively strong bound of $M_0 > 510\GeV$ on the effective mass $M_0$ defined in \Eqref{eq:eff-mass}. However, there is an error in the calculation. The final result should not depend on the Kaon mass $m_K$ but $m_\mu$ and the corrected expression in our model reads
\begin{equation}
\frac{\Gamma(K^0_L\to\mu^\pm e^\mp)}{\Gamma(K^+\to \mu^+\nu)}=\frac{9m_\mu^2 m_s^2}{|V_{us}|^2}\left[\frac{\sin^2\alpha}{m_{\Phi_1}^2}+\frac{\cos^2\alpha}{m_{\Phi_2}^2}\right]^2\;.
\end{equation}
The branching fraction is constrained to be less than $4.7\cdot 10^{-12}$. Using $m_s=95\,\MeV$, $m_\mu=106\MeV$, $V_{us}=0.225$, this leads to a bound of 
\begin{equation}
\frac{m_{\Phi_1} m_{\Phi_2}}{\sqrt{m_{\Phi_1}^2\cos^2\alpha+m_{\Phi_2}^2\sin^2\alpha}}\gtrsim 248\GeV\;.
\end{equation}

\mathversion{bold}
\subsection*{Quarks Transforming under generator $X$}
\mathversion{normal}
Another interesting possibility that is not possible in $A_4$ models is to assign the quarks to representations that also transform under the group generator $X$. Since the top mass is large, we want it to be generated at the renormalizable level, while all the other quark masses might well be the result of higher order effects. Looking at the multiplication rule 
\begin{align}
\MoreRep{3}{i}\times\MoreRep{3}{j}&=\sum_{\stackrel{k=1}{k\neq i,j}}^{5}\MoreRep{3}{k}, \hspace{2cm} (i\neq j),
\end{align}
it is clear that if one assigns $Q\sim (\MoreRep{3}{2},1)$ and $U^c\sim (\MoreRep{3}{3},1)$ there is only one Yukawa coupling at the renormalizable level
\begin{equation}
-\mathcal{L}_t= y_t\, Q \chi U^c +\hc\;,
\label{eq:top-mass-oper}
\end{equation}
which generates the top mass. The charm and up quark masses, as well as up sector mixing are generated by operators of the form
\begin{equation}
-\mathcal{L}_ u= y_i^{(u,1)}\,\left[ Q \chi U^c(\phi_1 \phi_1) \right]_i+y_i^{(u,2)}\,\left[ Q \chi U^c(\phi_2 \phi_2) \right]_i +\hc\;,
\end{equation}
where the sum goes over all singlet contractions of the fields. There are certainly enough parameters to fit the quark masses and up-type mixing. Actually, there are no further predictions besides the large top mass, since there are too many free parameters. 

In the down-type sector we can either utilize the same structure as in the up-type sector or, as the bottom quark mass is closer to the charm mass than to the top mass, we can use the assignment $D^c\sim (\MoreRep{3}{1},1) $. With this choice there is no tree-level operator of type \eqref{eq:top-mass-oper} allowed and all down type quark masses and mixing arise from 
\begin{equation}
-\mathcal{L}_ u= y_i^{(d,1)}\,\left[ Q \chi D^c(\phi_1 \phi_1) \right]_i+y_i^{(d,2)}\,\left[ Q \chi D^c(\phi_2 \phi_2) \right]_i +\hc\;
\end{equation}
We will not discuss this possibility further here, as we are primarily focused on the lepton sector.

\mathversion{bold}
\subsection*{Additional EW Higgs Doublet $H_q\sim\MoreRep{1}{1}$}
\mathversion{normal}
Another possibility is that the flavor structure in the quark sector could be completely unrelated to the one in the lepton sector. In particular, the quarks might not transform under the flavor symmetry in the lepton sector. This can be simply achieved by assigning the quarks to the singlet representation of the flavor group. In order to generate the quark mass matrices, we have to introduce an additional EW Higgs Doublet $H_q$, which does not transform under the flavor group. Hence, the flavor structure in the quark sector is unchanged compared to the SM one. Therefore, we do not discuss this possibility further and we will only briefly comment on its collider phenomenology in \Secref{sec:collider}.

The only effect\footnote{Here we assume that $H_q$ does not give a leading order contribution to the Weinberg operator. Symmetries can always be adjusted in order for this to be the case. If $H_q$ does give such a contribution there will be one more free physical phase in the neutrino mass matrix that cannot be rotated away.} of the additional Higgs doublet on the discussion in the preceding sections is to rescale the VEV of $H$ such that
$$
\VEV{H^0}^2+\VEV{H_q^0}^2=\frac{1}{2}(\sqrt{2}G_F)^{-1}=\frac{1}{2}(246 \GeV)^2
$$
is maintained.

%%%%%%%%%%%%%%%%%%%%%%%%%%%%%%%%%%%%%%%%%%%%%%%%%%%%%%%%%%%%%%%%%%%55
%%%%%%%%%%%%%%%%%%%%%%%%%%%%%%%%%%%%%%%%%%%%%%%%%%%%%%%%%%%%%%%%%%%55
%%%%%%%%%%%%%%%%%%%%%%%%%%%%%%%%%%%%%%%%%%%%%%%%%%%%%%%%%%%%%%%%%%%55
%\input{collider-v1}

\section{Collider Phenomenology\label{sec:collider}}
Our model predicts several new particles with EW charges at the EW scale. In this section, we will concentrate on the simplest extension to the quark sector given in \Secref{sec:quarks-simplest}, where quark doublets are assigned to the triplet representation $\MoreRep{3}{1}$ of the flavor group and obtain their masses from a coupling to the flavored Higgs $\chi$, as discussed in the previous section. We will briefly comment on the possibility to have a separate Higgs for the quark sector in the \Secref{sec:ColliderHq}.
Besides the fermionic singlets $S$, there are several EW doublets, which can be grouped in three different categories, the Higgs $h$, which obtains a VEV, the two partners of the Higgs in the flavor triplet $\chi$, namely $\varphi^\prime$ and $\varphi^{\prime\prime}$, and the additional inert EW scalar doublets $\hat\eta$. In the following, we sketch the different production and decay channels and discuss their implications for direct searches at colliders as well as the current bounds on the existence of new particles beyond the SM. However, a detailed study is beyond the scope of this presentation.

After a brief discussion of electroweak precision constraints and a short summary of the main experimental results, we will discuss each class of new particles separately.

\addtocontents{toc}{\protect\setcounter{tocdepth}{1}}
\subsection{Electroweak Precision Constraints}
\addtocontents{toc}{\protect\setcounter{tocdepth}{2}}

The experimentally measured values of the oblique parameters $S$ and $T$ have been obtained by several precision measurements at LEP and Tevatron.
The PDG~\cite{PDG:2012} quotes values of $S=-0.04\pm0.09$ and $T=0.07\pm 0.08$ at $95\%$ C.L. with a correlation between $S$ and $T$ of 88$\%$ for a reference value of $m_{h,ref}=117$ GeV. 

A general discussion in a multi-Higgs doublet model with an arbitrary number of Higgs doublets with hypercharge $Y=\pm\frac12$ and an arbitrary number of SM singlets has been given in \cite{Grimus:2008nb}. The expressions for the oblique parameters can be directly applied to this model, since the flavor symmetry only leads to additional restrictions on the masses and mixing matrices. We only estimate the contribution to $S$ and $T$ in the limit of small mixing in of $\hat\eta$, $\phi_i$, $H$ and only consider the mixing of $\varphi^\prime$ with $\varphi^{\prime\prime}$, which exactly corresponds to the region in parameter space being studied in the previous sections. In this limit also the charged and neutral scalar masses of the doublets $\hat\eta$ coincide. In this approximation, the contribution of $H$ exactly cancels with the subtracted SM term, the contribution of $\hat\eta$ and $\phi_i$ to $T$ vanishes, and $\phi_i$ does not contribute to $S$, since it does not couple to the gauge bosons in this approximation. Hence, the final contribution to $S$ originates from $\hat\eta$ and is given by
\begin{align}
S_{\eta,\phi}&=\frac{\cos^2 (2 \theta_W)}{24 \pi}\sum_{a=1}^n  \tilde G\left(\frac{m_Z^2}{m_a^2}\right) 
\end{align}
where $a=1,\dots n$ sums over the EW doublets contained in $\hat\eta$ with the charged scalar masses $m_a$ and $\theta_W$ denotes the Weinberg angle. 
The function $\tilde G$ is defined by
\begin{equation}
\tilde G(x)=-\frac{16}{3}+\frac{16}{z}-2 \left(\frac{4-z}{z}\right)^{3/2} \arctan \left(\frac{\sqrt{z(4-z)}}{2-z}\right)
\end{equation}
Its absolute value is monotonously decreasing for $z\to 0$ starting from $\tilde G(1)=-0.216$ to $\tilde G(0)=0$.
The contribution from $\varphi^{\prime(\prime)}$ to $S$ and $T$ is given by 
\begin{align}
T_{\varphi^{\prime(\prime)}}&=\frac{1}{8\pi \sin\theta_W^2 m_W^2} \sum_{a=1}^2\sum_{b=1}^2 \left|U_{\varphi,ba}\right|^2 F(m_a^2,\mu_b^2)\\
S_{\varphi^{\prime(\prime)}}&=\frac{1}{24 \pi} \sum_{a=1}^2  \left[
\cos^2 (2\theta_W) \tilde G\left(\frac{m_Z^2}{m_a^2}\right) 
+2\,\ln \frac{\mu_a^2}{m_a^2}
\right]
\end{align}
where the mixing matrix in the neutral $\varphi^{\prime(\prime)}$ sector, $U_\varphi$, is defined in \Eqref{eq:mixVarphiNeutral} and $m_a$ ($\mu_b$) denotes the charged (neutral) masses of the fields contained in $\varphi^{\prime(\prime)}$. The function $F$ is defined by
\begin{equation}
F(x,y)=\left\{\begin{array}{ccc}
\frac{x+y}{2}-\frac{xy}{x-y}\ln\frac{x}{y} & \mathrm{if} & x\neq y\\
0 & \mathrm{if} & x=y\\
\end{array}\right.\;.
\end{equation}
The next-to leading order corrections are suppressed by small mixing angles in the scalar sector. Hence, the model is consistent with electroweak precision tests in the phenomenologically interesting region, i.e. for small mixing in the scalar sector.

\addtocontents{toc}{\protect\setcounter{tocdepth}{1}}
\subsection{Summary of Relevant Experimental Results from Colliders}
\addtocontents{toc}{\protect\setcounter{tocdepth}{2}}

Recently, after the initial announcement of a SM-Higgs like resonance by ATLAS~\cite{:2012gk} and CMS~\cite{:2012gu}, which was mainly based on the diphoton as well as the $h\rightarrow ZZ^*\to 4l$ channel, several other channels have been measured or updated~\cite{ATLAS-CONF-2012-162,CMS-HIG-12-045}. We will briefly summarize the current status. The current best fit values for the mass of the resonance are $126.0 \pm 0.4(\mathrm{stat.}) \pm 0.4(\mathrm{sys.}) \GeV$ by ATLAS~\cite{:2012gk,ATLAS-CONF-2012-162} and  $125.8\pm 0.4(\mathrm{stat.}) \pm 0.4(\mathrm{sys.}) \GeV$ by CMS~\cite{CMS-HIG-12-045}.
The results are usually reported in terms of the signal strength normalized to the SM prediction, i.e.
\begin{equation*}
R_{X}\equiv \frac{\sigma(pp\to h) \mathrm{Br}(h\to X)}{\sigma(pp\to h_\mathrm{SM})\mathrm{Br}(h_\mathrm{SM}\to X)}\;.
\end{equation*}
The two main channels are the decay into two photons and $h\rightarrow ZZ^*\rightarrow 4 l$. While the $h\rightarrow ZZ^*\rightarrow 4l$ rate seems to agree with the SM prediction with $R_{ZZ}=1.2\pm0.6$ for ATLAS~\cite{ATLAS-CONF-2012-162} and $R_{ZZ}=0.8^{+0.35}_{-0.28}$ for CMS~\cite{CMS-HIG-12-045}, the $h\rightarrow \gamma \gamma$ rate seems to be enhanced with $R_{\gamma\gamma}=1.8\pm 0.5$ for ATLAS and $R_{\gamma\gamma}=1.56\pm0.43$ for CMS. 
The remaining channels include $h\to W W^*$ with a signal strength of $R_{WW}=1.4\pm0.6$ (ATLAS) and $R_{WW}=0.74\pm0.25$ (CMS) and the two channels with decays into fermions $hV\to b\bar b V$ with a signal strength of $R_{b\bar bV}=-0.4\pm 1.1$ (ATLAS) and $R_{b\bar bV}=1.3^{+0.7}_{-0.6}$ (CMS) as well as $h\to \tau\tau$ with $R_{\tau\bar\tau jj}=0.7\pm0.7$ (ATLAS) and $R_{\tau\bar\tau jj}=0.72\pm0.52$ (CMS).
All channels but the decay of the Higgs boson to two photons are in agreement with the SM prediction. The deviation in the diphoton channel is intriguing, as in the SM this decay proceeds via a loop diagram and is thus sensitive to new physics contributions. However, so far, the deviation is at the $1-2\sigma$ level~\cite{Baglio:2012et,Plehn:2012iz,ATLAS-CONF-2012-127}, if the uncertainties are taken into account conservatively.
Besides the discovery of a Higgs-like resonance, the LHC has put strong constraints on any physics beyond the SM.

Charged Higgs particles are constrained by searches at LEP and LHC. At LEP, charged Higgs particles $H^\pm$ are produced via a virtual $Z^*$ in the s-channel, i.e. $e^+e^-\rightarrow Z^* \rightarrow H^+ H^-$, and studied via their decays into $\tau \nu_\tau$ as well as $\bar c  s$ assuming their branching ratios add up to $1$, i.e. Br( $H^+ \rightarrow  \tau^+ \nu_\tau$ )+Br( $H^+ \rightarrow c \bar s$ )=1. This results in a bound of $m_{H^+}> 79.3\GeV$~\cite{PDG:2012}. Independent of any assumptions on the branching ratio, the invisible Z decay leads to $m_{H^+}\gtrsim 45\GeV$~\cite{PDG:2012}.
CMS searched for charged Higgs particles~\cite{CMS:2012cw}, which are produced in top decays, $t\rightarrow H^+ b$ and constrains their branching ratio Br($t\rightarrow H^+ b$) to less than 2\%-4\% for charged Higgs masses between $80$ and $160\GeV$. Similarly, the search by the ATLAS experiment~\cite{Aad:2012tj} yields bounds on the branching ratio Br($t\rightarrow H^+ b$) of the order of 1\%-5\% for charged Higgs masses in the range between $90$ and $160\GeV$, assuming Br($H^+\rightarrow \tau^+\nu_\tau$)=1.

\addtocontents{toc}{\protect\setcounter{tocdepth}{1}}
\subsection{EW Higgs Doublet $H$}
\addtocontents{toc}{\protect\setcounter{tocdepth}{2}}
\label{sec:Higgs}
We will first consider the limit in which there is no mixing between the Higgs $h$ and the flavons $\phi_i$. 
In the limit of no mixing, the tree-level couplings of the Higgs $h$ contained in the EW Higgs doublet $H$ to gauge bosons are identical to the SM couplings. In addition, the flavor conserving tree level couplings of the Higgs $h$ to fermions also agree with the SM ones. Note that there might be small corrections, since quark mixing vanishes at leading order and the Higgs couplings conserve all flavor numbers separately. As there are no new colored particles and the coupling of the Higgs to $t\bar t$ is the same as in the SM, the loop-induced coupling of the Higgs $h$ to gluons agrees with the SM one. In summary, the production of the Higgs $h$ as well as all tree-level decay channels and the decay into gluons are exactly like those in the SM. The Higgs decay into two photons is the only decay channel that can show a significant deviation from the SM in this approximation. If any of the other new scalars were light enough, there would be additional tree level Higgs decays into pairs of these scalars and such scenarios are therefore constrained. The decay $h\to S \bar S$ , if kinematically allowed, is loop suppressed.

Mixing of the Higgs $h$ with the flavons $\phi_i$ leads to a suppression of all tree-level couplings to gauge bosons and fermions. Hence, the production cross section is reduced according to the admixture of the flavons $\phi_i$ to the Higgs boson. As Higgs decays into $Z Z^*$ are close to the SM value, the admixture of the flavons $\phi_i$ to the Higgs $h$ is limited.

\begin{figure}\centering
\begin{minipage}{0.48\linewidth}
\includegraphics[width=\linewidth]{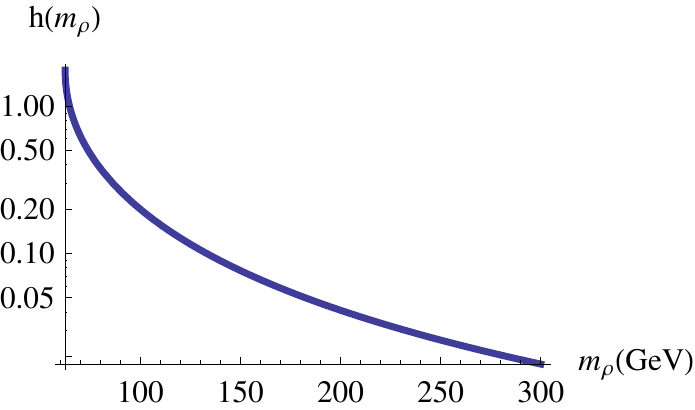}
\caption{Plot of function $h$ of Eq.~(\ref{eq:hfunct}).\label{fig:h-plot}}
\end{minipage}
\hfill
\begin{minipage}{0.48\linewidth}
\includegraphics[width=\linewidth]{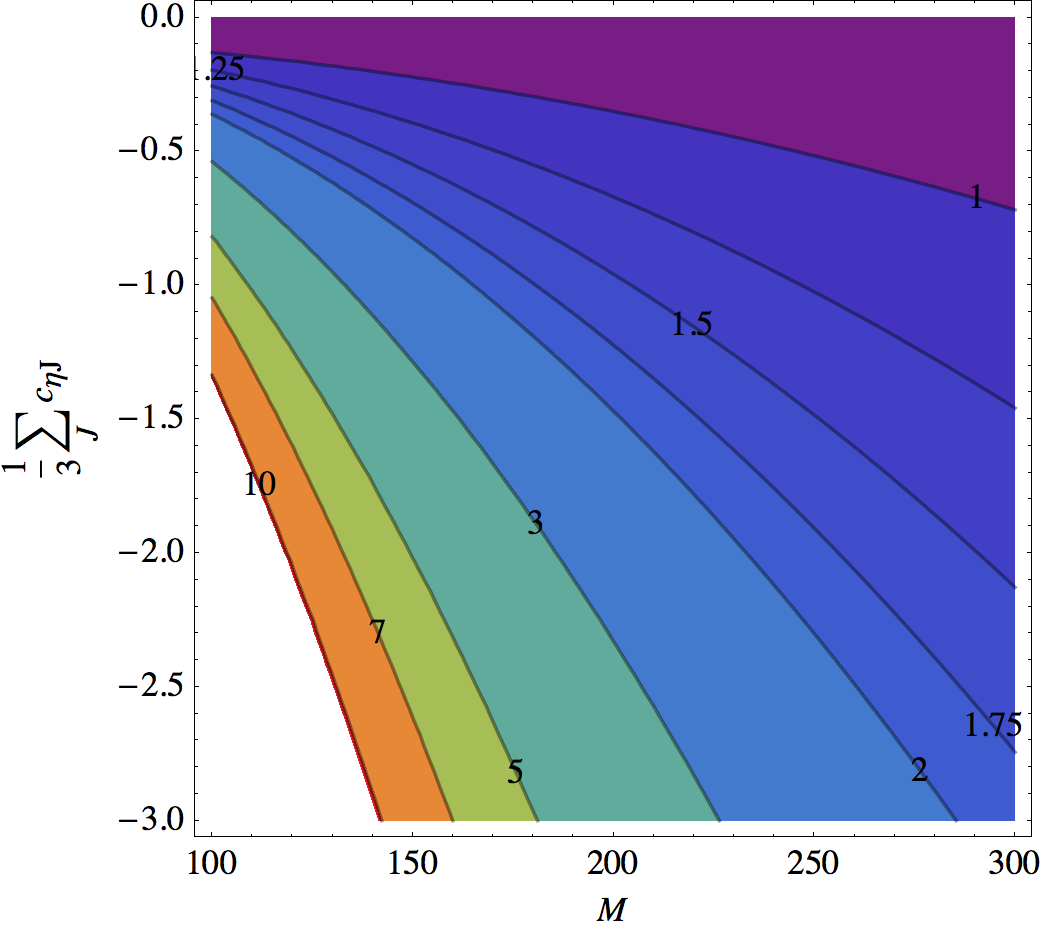}
\caption{$R_{\gamma \gamma}$ in the case where all charged scalars have the same common mass $M$ as a function of $\frac{1}{3}\sum_{J=1}^3 c_{\eta J}$.  \label{fig:sumJs-plot}}   
\end{minipage}
\end{figure}

Finally, let us discuss the diphoton decay channel. The SM contribution is dominated by the $W$ boson contribution and the smaller top loop contribution, which interfere destructively. In our model, the decay into two photons receives additional contributions from charged scalars in the loop, which are contained in the EW doublets $\varphi^\prime$, $\varphi^{\prime\prime}$ as well as $\hat\eta$. 
Any enhancing contribution has to interfere constructively with the SM $W$ boson loop or dominate over the $W$ boson contribution. The contribution of additional charged scalars $\rho_i$ with a charge one, coupled to the Higgs boson via the Higgs portal
\begin{align}
\mathcal{O}_{\rho_i}=c_{\rho_i} H^\dagger H \vert \rho_i\vert^2 \;,
\end{align}
has recently been studied in~\cite{Carena:2012xa}. The ratio of the effective coupling of the Higgs boson to two photons vs. the SM prediction is given by
\begin{align}
R_{\gamma \gamma}=\left \vert 1-\sum_{i} c_{\rho_i}h(m_{\rho_i})\right \vert^2\;,
\end{align}
where the function $h$ is depicted in \Figref{fig:h-plot}.
To obtain an enhancement of a factor of 2 (1.5), one thus needs a value of
\begin{align}
\sum_{i} c_{\rho_i}h(m_{\rho_i})=\left\lbrace\begin{array}{ccc}-0.41&  (-0.22)& \mathrm{for\ constructive\ interference } \\2.41 & (2.22)&\mathrm{for\ destructive\ interference } 
\end{array}\right\rbrace\;.
\label{eq:hfunct}
\end{align} 
Hence, a large negative coupling $c_\rho\sim -2$ is necessary to obtain an enhancement factor of 2 for a single singly charged scalar of mass $100 \GeV$.  Such a large negative coupling destabilizes the vacuum and leads to charge breaking minima unless $\vert c_\rho\vert< \sqrt{\lambda \lambda_\rho}\sim\sqrt{\lambda_\rho}/2$ is fulfilled, where $\lambda_\rho$ denotes the quartic coupling $\lambda_\rho\vert \rho \vert^4/2$. Note that this requires very large values for $\lambda_\rho$.   

Let us now use this formula to estimate the deviations from $R_{\gamma \gamma}=1$ that can be expected in this model. In total we have 11 charged scalars, 9 from the doublets $\hat\eta$ and two from the doublets $\varphi^\prime, \varphi^{\prime \prime}$. The interaction of the last two scalars with the Higgs field can be expressed as
\begin{align}
c_{\varphi^{\prime}}=\frac{m_h^2+m^2_{\varphi^{\prime+}} }{v^2},\qquad c_{\varphi^{\prime\prime}}=\frac{m_h^2+m^2_{\varphi^{\prime \prime+}} }{v^2},
\end{align}
with $m^2_{\varphi^{\prime (\prime)+}} $ defined in Eq.~(\ref{eq:masses-charged}). In the limit of large $m_{\varphi^{\prime(\prime)}}^2$ these two fields contribute 
\begin{align}
c_{\varphi^{\prime}} h({M^c_{+}})+c_{\varphi^{\prime\prime}} h({M^c_{-}})=0.1 +\left(\frac{22 \GeV}{{M^c_{+}}}\right)^2+\left(\frac{22 \GeV}{{M^c_{-}}}\right)^2\;.
\end{align}
The couplings of the charged components of the $\eta$ fields are given by
\begin{align}
\mathcal{O}_{\eta}=\sum_{i,J=1}^{3} c_{\eta_J} H^\dagger H \vert \eta_{J}^{(i)}\vert^2,
\end{align}
as dictated by the symmetry. The coefficients $c_{\eta_J}$ are essentially unconstrained except for the fact that the combination that couples to the DM particle should not be too large, to avoid the bound from direct detection. In the limit where all charged scalars have a common mass $M$, we see from \Figref{fig:sumJs-plot} that $M=200 \GeV$ requires $\frac{1}{3}\sum_{J=1}^{3}c_{\eta_J} =-1.46$ ($-0.95$) for $R_{\gamma \gamma}=2(1.5)$. 

In case the $h\rightarrow \gamma \gamma$ anomaly persists, it would be interesting to measure $h\rightarrow \gamma Z$, since it originates from similar diagrams, where one photon is replaced by one $Z$ boson. A cross-correlation of the two measurements would allow one to determine the isospin of these particles. In our model all charged scalars are part of $\SU{2}$ doublets allowing us to distinguish it from other models which have EW multiplets in the loop with different EW charges, like singlets or triplets.

\addtocontents{toc}{\protect\setcounter{tocdepth}{1}}
\subsection{Further Scalars}
\addtocontents{toc}{\protect\setcounter{tocdepth}{2}}
Besides the Higgs $h$, there are several additional scalars, such as the flavor-violating EW scalar doublets $\varphi^{\prime(\prime)}$ as well as $\hat\eta$, which do not acquire a VEV, and the flavons $\phi_i$, which acquire a VEV. See \Appref{sec:vac-align-and-spectrum} for the scalar mass spectrum.

\subsubsection{Flavor-violating Higgs Doublets $\varphi^{\prime(\prime)}$}
The neutral components of the flavor-violating Higgs doublets $\varphi^{\prime(\prime)}$ have neither tree-level couplings to $t\bar t$ nor couple to two EW gauge bosons at tree level. Hence, they are not produced by any of the standard Higgs production channels, but they can be produced via associate production with two different quarks, $gg\to  \varphi^{\prime(\prime)} q q^\prime$. The dominant channel is $gg\to  \varphi^{\prime} \bar t u, \varphi^{\prime\prime} \bar t c$, which has a cross section of the same magnitude as production of a Higgs boson in association with a $t \bar t$ pair. Therefore, they are not constrained by the current heavy Higgs searches. Other productions channels are $q q^\prime\to \varphi^{\prime(\prime)}$, where $q$ and $q^\prime$ are in different generations as well as pair-production in vector boson fusion $WW,ZZ\to \varphi^\prime \varphi^\prime,\varphi^{\prime\prime} \varphi^{\prime\prime}$. These processes are suppressed compared to the main Higgs production channels at the LHC. Note, however, that the decays of the flavor-violating Higgs doublets $\varphi^{\prime(\prime)}$ might lead to distinct flavor-violating signatures similar to the recent analyses of flavor-violating Higgs decays in models with flavor symmetries~\cite{Bhattacharyya:2010hp,Cao:2010mp,Cao:2011df,Bhattacharyya:2012ze,Davidson:2012ds}.

As the charged Higgs particles contained in $\varphi^{\prime(\prime)}$ do not couple to $t\bar b$, the LHC limits do not apply. Hence, the charged Higgs particles in our model are only constrained by the LEP limits discussed previously.

Although there are no constraints yet, upcoming searches will test the allowed range of masses, because the flavor-violating Higgs doublets $\varphi^{\prime(\prime)}$ stem from the same flavor triplet as the Higgs doublet $H$, and therefore their masses are determined to be given by scalar couplings times the EW VEV. Their masses may therefore not be raised arbitrarily high, as discussed below Eq.~\eqref{eq:masses-charged}\footnote{Note that if one introduces soft-breaking terms that respect the $Z_3$ symmetry, it is possible to adjust the mass terms arbitrarily~\cite{Ma:2010gs}. Alternatively one may introduce an EW singlet scalar that transforms as $\MoreRep{3}{1}$ and breaks to the same subgroup as $\chi$. This can be  realized without introducing a vacuum alignment problem.}.

\mathversion{bold}
\subsubsection{EW Scalar Doublets $\hat\eta$}
\mathversion{normal}
The neutral components of the EW scalar doublets $\hat\eta$ do not couple to quarks and particularly not to $t\bar t$ as well as two EW gauge bosons. They can be pair-produced in vector boson fusion $WW,ZZ\to \eta_i\eta_i$. Hence, similarly to the flavor-violating Higgs doublets, they are not produced via the main Higgs production channels and the current bounds from heavy Higgs searches do not constrain $\hat\eta$. Also, the charged components of $\hat\eta$ are not constrained by the current LHC searches, because they do not couple to quarks directly and the charged Higgs bounds do not apply. Therefore, they are only constrained by the LEP searches.

\mathversion{bold}
\subsubsection{Flavons $\phi_i$}
\mathversion{normal}
The flavons $\phi_i$ do not have gauge interactions and they do not couple to fermions directly. However, they mix with the Higgs $h$, which is constrained by the Higgs searches to be small, since a large mixing suppresses the production cross section of $h$ and therefore all rates relative to the SM expectation. In conclusion, the scalar mass eigenstates which are dominantly composed of the flavons $\phi_i$ are only produced via mixing with the Higgs $h$ and thus there are no limits from current searches due small mixing. 

\mathversion{bold}
\subsection{Variant with Additional EW Higgs Doublet $H_q$}
\label{sec:ColliderHq}
\mathversion{normal}
As we discussed in \Secref{sec:quarks}, another simple possibility to incorporate quarks in the model is by assigning all quarks to the trivial representation of the flavor group and introducing an additional EW Higgs doublet $H_q$, which transforms trivially under the flavor group. This leads to different collider signatures compared to the previously discussed scenario. Soon, these scenarios can be experimentally distinguished at the LHC. We will highlight the most important differences. 

The discussion of the fermions $S$ as well as the scalars $\hat\eta$ remains the same. 
The main changes are in the Higgs phenomenology. In contrast to the other scenario, the component in $\chi$ which obtains a VEV does not couple to quarks and therefore it is not produced in gluon fusion, unless there is mixing between $\chi$ and $H_q$. Instead, the newly introduced Higgs $H_q$ will be produced in gluon fusion. In this setup, the observed resonance at 125 GeV would be associated with the mass eigenstate, which is dominantly composed of Higgs $H_q$. As $H_q$ has exactly the same couplings to gauge bosons and quarks, but does not couple to leptons (especially $\tau$'s), the decays into leptons are suppressed by the mixing between $H_q$ and $H$ (contained in $\chi$). The diphoton branching ratio can be enhanced in the same way as discussed in \Secref{sec:Higgs}.

\addtocontents{toc}{\protect\setcounter{tocdepth}{1}}
\subsection{Fermionic Singlets $S$}
\addtocontents{toc}{\protect\setcounter{tocdepth}{2}}
The additional fermionic states $S$ are SM singlets and only charged under the discrete flavor group. Furthermore, they only couple to lepton doublets and therefore their production cross section at hadron colliders is suppressed compared to colored particles and there are no relevant analyses at present. 
The production depends on the exact mass spectrum of $\hat\eta$ as well as $S$. 
The production via t-channel $\hat\eta$ exchange is always present in a lepton collider, e.g. $e^+ e^-\rightarrow S \bar S$.
If $S$ is lighter than one of the components of $\hat\eta$, it is possible to produce $S$ via EW production of these heavier components of $\hat\eta$ and subsequently decay into $S$ and one lepton.
Unless $S$ is the DM candidate, the fermionic singlet $S$ will decay into a lepton and one of the lighter components of $\hat\eta$, which will subsequently cascade down to DM via EW gauge interactions.
The signal is missing transverse energy and leptons (and possibly EW gauge bosons) in the final state.

If $S$ is lighter than all components of $\hat\eta$ and therefore a DM candidate, there are bounds from mono-photon searches at LEP~\cite{Fox:2011fx}. As $S$ only couples to leptons, the searches at hadron colliders are weaker due to the additional suppression from loops that couple leptons to quarks.
The mono-photon searches at LEP probe the effective DM annihilation operator $(\bar e S) (e\bar S)/\Lambda_t^2$, which are induced by the exchange of a scalar doublet $\eta_{1,2}$. The scale $\Lambda_t$ of this operator is determined by $\Lambda_t^{-2}=\sum_k |h_k|^2/M_k^2$ for $M_k\gg M_S$. The analysis in~\cite{Fox:2011fx} quotes a limit of $(200-340) \GeV$ for $M_S<90\GeV$. Hence, this does not impose a strong constraint, since the smallness of neutrino masses points towards larger cutoff scales $\Lambda_t$.

%%%%%%%%%%%%%%%%%%%%%%%%%%%%%%%%%%%%%%%%%%%%%%%%%%%%%%%%%%%%%%%%%%%55
%%%%%%%%%%%%%%%%%%%%%%%%%%%%%%%%%%%%%%%%%%%%%%%%%%%%%%%%%%%%%%%%%%%55
%%%%%%%%%%%%%%%%%%%%%%%%%%%%%%%%%%%%%%%%%%%%%%%%%%%%%%%%%%%%%%%%%%%55

%\input{Conclusions}
\section{Conclusions}
\label{sec:conclusions}

We presented a predictive renormalizable $A_4$ model of lepton flavor at the electroweak scale. The flavor group $A_4$ is extended in the scalar potential to $Q_8\rtimes A_4$, which allows a natural vacuum alignment at the EW scale~\cite{Holthausen:2011vd}. This is the first model of its kind that explains the lepton flavor structure at the EW scale including the correct vacuum alignment.

The SM Higgs boson is subsumed in a flavor triplet that couples to charged leptons (and quarks) at the renormalizable level, thereby eliminating the need to invoke higher dimensional operators, as is done in models with flavon singlets. 
Neutrino masses are generated at the one-loop level and are further suppressed by the fact that two small mass insertions are needed in the loop. 
This TeV seesaw is  realized without imposing any new symmetries apart from the flavor symmetries. 
In the model there are five real free parameters, which gives a predictive framework and, in particular, a correlation between the atmospheric and reactor angle is predicted, which agrees well with the recent global fits. 
Furthermore the model automatically includes a WIMP dark matter candidate and its stability and phenomenology have been studied. It can explain the observed dark matter abundance and is consistent with current exclusion limits by dark matter detection experiments.
Constraints from LFV experiments are loosened by the flavor symmetry in comparison to flavor generic multi-Higgs doublet models due to the remnant $Z_3$ symmetry in the charged lepton sector. 

Finally, several possible extensions to the quark sector have been studied. We studied the collider phenomenology of the simplest extension to the quark sector, which does not require the introduction of new particles at leading order and in which the quarks multiplets transform like the lepton multiplets under the flavor symmetry, and commented on the other possibilities. We studied the possibility of the Higgs boson $h$ to explain the observed resonance at $125 \GeV$, especially the enhanced diphoton rate, which can be straightforwardly explained by the multitude of additional charged particles contained in the EW scalar doublets, which all contribute to the radiative decay of $h\to\gamma\gamma$. 
The fact that the Higgs doublet is contained in a flavor triplet leads to distinct signatures at the LHC. There are additional EW scalar doublets $\varphi^{\prime(\prime)}$, which cannot be decoupled from the Higgs $h$, and therefore are accessible in searches at the LHC. As they do not acquire a VEV, they do not decay into gauge bosons, but only via Yukawa type interactions into fermions besides decays into other scalars. Because of the remnant $Z_3$ flavor symmetry in the charged lepton sector, they only exhibit flavor violating decays into fermions in contrast to the Higgs $h$.

It might be interesting to study leptogenesis in this model. Because of the flavor symmetry, the fermionic SM singlet $S$ are degenerate in mass at tree level as well as all couplings but $\lambda_2$ are real. The degeneracy is only lifted at two loop order and therefore the induced mass splittings are small and there might be a resonant enhancement. This also introduces some CP violation into the mass matrix of $S$, but it has to be checked whether it is sufficient. We will leave a study of possible ways to obtain the baryon asymmetry of the Universe for future work.

%%%%%%%%%%%%%%%%%%%%%%%%%%%%%%%%%%%%%%%%%%%%%%%%%%%%%%%%%%%%%%%%%%%55
%%%%%%%%%%%%%%%%%%%%%%%%%%%%%%%%%%%%%%%%%%%%%%%%%%%%%%%%%%%%%%%%%%%55
%%%%%%%%%%%%%%%%%%%%%%%%%%%%%%%%%%%%%%%%%%%%%%%%%%%%%%%%%%%%%%%%%%%55

\appendix

\section*{Acknowledgements}
MH~acknowledges support by the International Max Planck Research School for Precision Tests of Fundamental Symmetries. ML and MS thank the Galileo Galilei Institute for Theoretical Physics for its hospitality.  MS would like to thank K.~Petraki for discussions and would like to acknowledge MPI f\"ur Kernphysik, where part of this work was done, for hospitality of its staff and the generous
support. This work was supported in part by the Australian Research Council.

%%%%%%%%%%%%%%%%%%%%%%%%%%%%%%%%%%%%%%%%%%%%%%%%%%%%%%%%%%%%%%%%%%%55
%%%%%%%%%%%%%%%%%%%%%%%%%%%%%%%%%%%%%%%%%%%%%%%%%%%%%%%%%%%%%%%%%%%55
%%%%%%%%%%%%%%%%%%%%%%%%%%%%%%%%%%%%%%%%%%%%%%%%%%%%%%%%%%%%%%%%%%%55

%\input{appendix-v1}
\section{Vacuum Alignment and Scalar Spectrum of }
\label{sec:vac-align-and-spectrum}
\subsection{Vacuum alignment} 
\label{sec:vacuum-alignment-app}
The vacuum configuration given in \Eqref{eq:vac-conf}
is naturally obtained from the most general potential~\footnote{We do not have to consider the part involving $\hat\eta$, because it does not change the minimization conditions of $\phi_i$ and $\chi$, if it does not acquire a VEV.} 
$$
V=V_\phi(\phi_1,\phi_2)+V_\chi(\chi)+V_{\mathrm{mix}}(\chi,\phi_1,\phi_2)
$$
compatible with given symmetries, where  $V_\chi(\chi)$ is given in \eqref{eq:chi-potential} and 
\begin{align}
V_\phi(\phi_1,\phi_2)=& \mu_1^2 (\phi_1 \phi_1)_{\MoreRep{1}{1}}+\alpha_1 (\phi_1 \phi_1)_{\MoreRep{1}{1}}^2 + \sum_{i=2,3}\alpha_i (\phi_1 \phi_1) _{\MoreRep{3}{i}} \cdot (\phi_1 \phi_1) _{\MoreRep{3}{i}}\nonumber \\
+&\mu_2^2 (\phi_2 \phi_2)_{\MoreRep{1}{1}}+\beta_1 (\phi_2 \phi_2)_{\MoreRep{1}{1}}^2 + \sum_{i=2,3}\beta_i (\phi_2 \phi_2) _{\MoreRep{3}{i}} \cdot (\phi_2 \phi_2) _{\MoreRep{3}{i}}\nonumber\\
+&\gamma_1 (\phi_1 \phi_1)_{\MoreRep{1}{1}} (\phi_2 \phi_2)_{\MoreRep{1}{1}}+ \sum_{i=2,3,4}\gamma_i (\phi_1 \phi_1) _{\MoreRep{3}{i}} \cdot (\phi_2 \phi_2) _{\MoreRep{3}{i}},\nonumber\\
V_\chi(\chi)&=\mu^2_3 (\chi \chi)_{\MoreRep{1}{1}}+\rho_1 (\chi \chi \chi)_{\MoreRep{1}{1}}+ \lambda_1 (\chi \chi)_{\MoreRep{1}{1}}^2+\lambda_2  (\chi \chi)_{\MoreRep{1}{2}}(\chi \chi)_{\MoreRep{1}{3}},
\nonumber\\
V_{\mathrm{mix}}(\chi,\phi_1,\phi_2)&=\zeta_{13} (\phi_1 \phi_1)_{\MoreRep{1}{1}} (\chi^\dagger \chi)_{\MoreRep{1}{1}} + \zeta_{23} (\phi_2 \phi_2)_{\MoreRep{1}{1}} (\chi^\dagger \chi)_{\MoreRep{1}{1}} 
\end{align}
compatible with given symmetries. The minimization conditions reduce to the equations
\begin{align}
a\left( \alpha _+ \left(a^2+b^2\right)+\alpha _- \left(a^2-b^2\right)+\gamma _+ \left(c^2+d^2\right)+\gamma _- \left(c^2-d^2\right)+U_1\right) +\Gamma b c d &=0\nonumber\\\nonumber
b\left(\alpha _+ \left(a^2+b^2\right)- \alpha _- \left(a^2-b^2\right)+\gamma _+ \left(c^2+d^2\right)-\gamma _- \left(c^2-d^2\right)+U_1\right) +\Gamma a c d&=0 \\
c\left(\beta _+ \left(c^2+d^2\right)+\beta _- \left(c^2-d^2\right)+\gamma _+ \left(a^2+b^2\right)+ \gamma _- \left(a^2-b^2\right)+U_2 \right) +\Gamma a b  d&=0\\
d\left(\beta _+ \left(c^2+d^2\right)-\beta _- \left(c^2-d^2\right)+\gamma _+ \left(a^2+b^2\right)- \gamma _- \left(a^2-b^2\right)+U_2 \right)+\Gamma a b c &=0 \nonumber\\\nonumber
 v\left( M_\chi^2+ \lambda_\chi v^2\right)&=0\
\end{align}
with 
\begin{align*}
U_i&=\frac{1}{2} \mu _i^2+\frac{\sqrt{3}}{12} \zeta _{i3}\, v^{2} \quad \mathrm{for}\quad i=1,2\;, 
\end{align*}
and
\begin{align*}
M_\chi^2 &=2 \mu_3^2+\zeta_{13}(a^2+b^2)+\zeta_{23} (c^2+d^2), \qquad 
\lambda_\chi =\frac{2}{3}\left(\sqrt{3} \lambda _{\chi \MoreRep{1}{1}}+\lambda _{\chi \MoreRep{3}{1S}}\right), \\
\xi_+&=\frac{\xi_1}{2},\quad \xi_-=\frac{\xi_2+\xi_3}{2\sqrt{3}},\qquad
\gamma_+=\frac{\sqrt{3}\gamma_1+\gamma_4}{4\sqrt{3}},\quad\gamma_-=\frac{\gamma_2+\gamma_3}{4\sqrt{3}},\quad\mathrm{and}\quad\Gamma=\frac{ \gamma_4}{\sqrt{3}},
\end{align*}
with $ \xi=\alpha,\beta$.  Since the number of equations matches the number of VEVs, vacuum alignment is possible. Corrections to the scalar potential only arise on dimension 6 level.
These corrections furthermore arise on one-loop level and are thus further suppressed. We therefore neglect VEV shifts arising from these 
interactions throughout this work.

\subsection{Scalar Spectrum }
\label{sec:scalar-spectrum}
\subsubsection{Scalar Spectrum -- $\phi_i$, $\chi$}
Let us first discuss the visible sector, i.e. the flavons $\phi_1,\phi_2,\chi$ that get VEVs and  realize the symmetry breaking; the $\eta$'s are independent and will be discussed later. 
The fields can be classified according to remnant symmetries of the potential. There are the obvious symmetries
\begin{align}
Z_3&: \chi\rightarrow T_3 \chi, \qquad \phi_i\rightarrow \phi_i,
\intertext{with $T_3=\Omega_T \diag (1, \omega^2,\omega)\Omega_T^\dagger$ and}
Z_2&: \phi_i\rightarrow S_4 \phi_i, \qquad \chi\rightarrow \chi,
\intertext{with $S_4=\Omega_{S_4} \diag (1, 1,-1,-1)\Omega_{S_4}^\dagger$ but there is another accidental symmetry of the potential $V_\phi$ not part of $Q_8 \rtimes A_4$:}
Z_2&: \phi_i\rightarrow O_4 \phi_i, \qquad \chi\rightarrow \chi,
\end{align}
with\footnote{The alert reader will recognize this as an outer automorphism $h_4$ defined in \cite{Holthausen:2012dk}.}  $O_4=\Omega_{S_4} \diag (1, 1,1,-1)\Omega_{S_4}^\dagger$, where
\begin{align}
\Omega_{S_4}\equiv\frac{1}{\sqrt{2}}\left(
\begin{array}{cccc}
 0 & 1 & 0 & -1 \\
 0 & 1 & 0 & 1 \\
 -1 & 0 & 1 & 0 \\
 1 & 0 & 1 & 0
\end{array}
\right).
\end{align} It is useful to go to a basis
\begin{align}
\tilde{\phi_i}=\Omega_{S_4}^\dagger \phi_i, \qquad \left(H, {\varphi^{\prime}}, {\varphi^{\prime \prime}} \right)^T=\Omega_T^\dagger  \chi, \qquad \left(L_e, L_\mu, L_\tau \right)^T=\Omega_T^\dagger L
\end{align}
where these symmetries are represented diagonally. Let us discuss the mass terms in turn:
\begin{itemize}
\item the 9 physical scalars contained in $\chi$ have been discussed following Eq.~\eqref{eq:masses-charged}
Here we only report the expressions of the dimensionless couplings in terms of masses:
\begin{align}
\lambda_{\chi 1_1}&={M^2_-}+{M^2_+}+\frac{3 {m_h^2}}{2} \nonumber\\
\lambda_{\chi 1_2}&=\frac{1}{2} \left(3 m_1^2-3 \sqrt{m_1^4-2 m_1^2 m_2^2+m_2^4-4
   \left({M^2_-}-{M^2_+}\right)^2}+3 m_2^2-2 {M^2_-}-2 {M^2_+}\right)\nonumber \\
\lambda_{\chi 3_1,S}&=-\sqrt{3} \left({M^2_-}+{M^2_+}\right)\nonumber \\
\lambda_{\chi 3_1,A}&= -\sqrt{3} \left(m_1^2+\sqrt{m_1^4-2 m_1^2 m_2^2+m_2^4-4
   \left({M^2_-}-{M^2_+}\right)^2}+m_2^2-{M^2_-}-{M^2_+}\right) \nonumber\\
\lambda_{\chi A}&= 6 \left({M^2_-}-{M^2_+}\right)
\end{align}

\item $(\tilde{\phi}_1)_4$ and $(\tilde{\phi}_2)_4$ transform as $(1,-1,-1)$ and have a mass matrix given by
$$
\left(
\begin{array}{cc}
m_{11} & \frac{2 \left(a c \left(\sqrt{3} \gamma_M-2
   \gamma_2\right)+2 b \gamma_2 d\right)}{\sqrt{3}} \\
.& m_{11}\left((a,b,c,d,\alpha_2)\leftrightarrow(c,d,a,b,\beta_2) \right)
\end{array}
\right)
$$
with 
\begin{align*}
m_{11}&= -4 \sqrt{3} a^2 \alpha_2+a \left(\frac{2 a \gamma_M (c-d) (c+d)}{(b-a)
   (a+b)}+\frac{c \Gamma  d}{b}\right)-\frac{b c \Gamma  d}{2 a}\nonumber\\&+\frac{1}{12} \left(48
   \sqrt{3} \alpha_2 b^2-3 \Gamma  \left(c^2+d^2\right)+8 \sqrt{3} \gamma_2 \left(d^2-c^2\right)\right)
\end{align*}
\item  $(\tilde{\phi}_1)_3$ and $(\tilde{\phi}_2)_3$ transform as $(1,-1,1)$ and have a mass matrix given by
$$
\left(
\begin{array}{cc}
m_{11} & \frac{2 \left(a c \gamma_2-b d \left(\gamma_2-2 \sqrt{3}
   \gamma_M\right)\right)}{\sqrt{3}} \\
.& m_{11}\left((a,b,c,d,\alpha_2)\leftrightarrow(c,d,a,b,\beta_2) \right)
\end{array}
\right)
$$
with 
\begin{align*}
m_{11}&=2 \sqrt{3} a^2 \alpha_2+\frac{2 b^2 \left(\sqrt{3} \alpha_2
   (a-b) (a+b)+2 \gamma_M (c-d) (c+d)\right)}{b^2-a^2}-\frac{a c \Gamma 
   d}{b}\\&+\frac{2 b c \Gamma  d}{a}-\frac{1}{2} \Gamma 
   \left(c^2+d^2\right)+\frac{\gamma_2 (c-d) (c+d)}{\sqrt{3}}
\end{align*}
\item the real scalars $h$, $(\tilde{\phi}_1)_1$, $(\tilde{\phi}_1)_2$, $(\tilde{\phi}_2)_1$ and $(\tilde{\phi}_2)_2$ transform as $(1,1,1)$ under the remnant symmetry. Here we don't give the full mass matrix but only give the mixing with the Higgs in the limit of small mixings. The mixing matrix with field $f$ is given by 
\begin{align}
\tan 2 \theta_f= \frac{2m_{h,f}}{m_f^2-m_h^2}
\label{eq:Higss-mixing-angles}
\end{align}
with 
\begin{align*}
m_{h,(\tilde{\phi}_1)_1}= -\frac{b v{\zeta_{13}}}{\sqrt{3}},\quad
m_{h,(\tilde{\phi}_1)_2}=  \frac{a v{\zeta_{13}}}{\sqrt{3}},\quad
m_{h,(\tilde{\phi}_2)_1}=  -\frac{d v{\zeta_{23}}}{\sqrt{3}},\quad
m_{h,(\tilde{\phi}_2)_2}=  \frac{c v{\zeta_{23}}}{\sqrt{3}}.
\end{align*}

\end{itemize}

\mathversion{bold}
\subsubsection{Scalar Spectrum -- $\hat\eta$}
\mathversion{normal}
The relevant part of the scalar potential to calculate the mass insertions needed to calculate neutrino masses  for the mass spectrum of $\hat\eta$ has been given in Eqs.~(\ref{eq:Veta2}-\ref{eq:Vetaphi}). To calculate the $\eta$ mass spectrum the complete interactions 
\begin{align}\nonumber
\delta V_{\hat\eta}^{(2)}&= \lambda_{1} (\chi^T \sigma_2 \vec{\sigma}  \chi)_{\MoreRep{1}{1}} (\eta_1^T \sigma_2 \vec{\sigma}  \eta_3)^*_{\MoreRep{1}{1}}
+\lambda_{2} e^{\I \alpha_{\lambda}}(\chi^T \sigma_2 \vec{\sigma}  \chi)_{\MoreRep{3}{1}} (\eta_2^T \sigma_2 \vec{\sigma}  \eta_3)^*_{\MoreRep{3}{1}} \\\label{eq:deltaVeta}
&+\lambda_3 (\phi_1 \phi_2)_{\MoreRep{1}{1}} (\eta_3^\dagger \eta_1)_{\MoreRep{1}{1}}
+\lambda_4(\phi_1 \phi_2)_{\MoreRep{3}{1}} (\eta_3^\dagger \eta_2)_{\MoreRep{3}{1}}
+\lambda_5(\phi_1 \phi_2)_{\MoreRep{3}{2}} (\eta_3^\dagger \eta_2)_{\MoreRep{3}{2}}\\\nonumber
&+\lambda_6(\phi_1 \phi_2)_{\MoreRep{3}{3}} (\eta_3^\dagger \eta_2)_{\MoreRep{3}{3}}
+\lambda_7(\phi_1 \phi_2)_{\MoreRep{3}{5}} (\eta_1^\dagger \eta_3)_{\MoreRep{3}{5,S}}
+\lambda_8(\phi_1 \phi_2)_{\MoreRep{3}{5}} (\eta_1^\dagger \eta_3)_{\MoreRep{3}{5,A}}\\\nonumber
&+l_{1}^{ij} (\phi_j \phi_j)_{\MoreRep{1}{1}} (\eta_i^\dagger \eta_i)_{\MoreRep{1}{1}}+
 l_{2}^{j}(\phi_j \phi_j)_{\MoreRep{3}{2,3}} (\eta_1^\dagger \eta_2)_{\MoreRep{3}{2,3}}+
l_{3}^{j}(\phi_j \phi_j)_{\MoreRep{3}{4}} (\eta_2^\dagger \eta_2)_{\MoreRep{3}{4}}\\\nonumber
&+k_1(\chi^\dagger \chi)_{\MoreRep{3}{1}} (\eta_1^\dagger \eta_2)_{\MoreRep{3}{1}} +k_2 (\chi^\dagger\sigma_2 \vec{\sigma} \chi)_{\MoreRep{3}{1}} (\eta_1^\dagger\sigma_2\vec{\sigma}  \eta_2)_{\MoreRep{3}{1}}\\\nonumber
&+k_3^{(i)}  (\chi^\dagger\sigma_2\vec{\sigma} \chi)_{\MoreRep{1}{1}} (\eta_i^\dagger \sigma_2\vec{\sigma}\eta_i)_{\MoreRep{1}{1}}+k_4^{(i)} (\chi^\dagger \chi)_{\MoreRep{1}{1}} (\eta_i^\dagger \eta_i)_{\MoreRep{1}{1}}+\hc
\end{align}
are needed. Let us briefly outline how the various couplings act: The couplings $k_4^{(i)}$ and $l_1^{(ij)}$ renormalize $M_i$, $k_3^{(i)}$ splits masses of charged and neutral components, $\lambda_1$ and $\lambda_2$  mix neutral scalar and pseudoscalar components of the various fields. Hence, it also splits the masses of scalar and pseudoscalar of the lightest mass eigenstate, $k_1$, $l_2^{(i)} $, $l_3^{(j)} $ mix the components of the various $\hat\eta$ and adds flavor breaking effects. Since $\vev{\chi^2_{\MoreRep{1}{2,3}}}=0$ such couplings do give contributions to mass terms and are not shown here. $\lambda_3,\dots, \lambda_8$ break $Z_4$ and therefore mix components of $\eta_3$ with components of $\eta_{1,2}$.

\section{Group Theory}

In this section, we give a short review of the relevant group theory of $Q_8\rtimes A_4$. 
We give the presentation of the group and a possible set of generators for all irreducible representations of the group. We summarize the most important Clebsch-Gordan coefficients for the quartet $\MoreRep{4}{1}$ and triplets \MoreRep{3}{i}. See \cite{Holthausen:2011vd} for a more detailed description of the group theory of $Q_8\rtimes A_4$. All Clebsch-Gordan coefficients can be obtained with the help of the Mathematica package \texttt{Discrete}, which has been published as part of \cite{Holthausen:2011vd}.

\subsection{Mini-Review}
The semidirect product $Q_8\rtimes A_4$ we are using is defined by the relations 
\begin{align}
SXS^{-1}&=X, & SYS^{-1}&=Y^{-1}, & % SYS^{-1}&=X^2Y
TXT^{-1}&=YX, & TYT^{-1}&=X\;.
\end{align}
between the generators of $A_4$
\begin{align}
\braket{S,T\vert S^2=T^3=(ST)^3=1}\;,
\end{align}
and $Q_8$
\begin{align}
\braket{X,Y\vert X^4=1,\; X^2=Y^2,\; Y^{-1}XY=X^{-1}}\;. %or XY=Y^3X
\end{align}
Note that it is sufficient to use e.g. the generators $X,S,T$ as $Y=T^{-1}XT$. The defining representation matrices for the representations we are using are given in \Tabref{tab:Q8rtimesA4representations} with 
\begin{align}
S_3&=\left(\begin{array}{ccc}
1&0&0\\
0&-1&0\\
0&0&-1
\end{array}\right)
& 
T_3&=\left(\begin{array}{ccc}
0&1&0\\
0&0&1\\
1&0&0
\end{array}\right)
& 
T_4&=\left(\begin{array}{cccc}
0&1&0&0\\
0&0&1&0\\
1&0&0&0\\
0&0&0&1
\end{array}\right)
\end{align}
and $
S_4=\sigma_3 \otimes\sigma_1$ and $X_4=-\I \sigma_2 \otimes\sigma_3$.

\begin{table}
\centering
\ra{1.2}
\begin{tabular}{lcccccccccccccc}\toprule
&&$\MoreRep{1}{1}$&$\MoreRep{1}{2}$&$\MoreRep{1}{3}$&&$\MoreRep{3}{1}$&$\MoreRep{3}{2}$&$\MoreRep{3}{3}$&$\MoreRep{3}{4}$&$\MoreRep{3}{5}$&&$\MoreRep{4}{1}$&$\MoreRep{4}{2}$&$\MoreRep{4}{3}$\\ \cmidrule{3-5}\cmidrule{7-11}\cmidrule{13-15}
$S$ &&$1$ &$1$ &$1$ && $S_3$&$T_3 S_3 T_3^2$ &$T_3 S_3 T_3^2$ & $\mathbbm{1}_3$ &  $T_3^2 S_3 T_3$ & & $S_4$ & $S_4$ & $S_4$ \\
$T$&&$1$&$\omega$&$\omega^2$&&$T_3$ &$T_3$ &$T_3$ &$T_3$ &$T_3$ && $ T_4$ & $\omega^2 T_4$ & $\omega T_4$ \\
$X$&&$1$&$1$&$1$&&$\mathbbm{1}_3$&$S_3$   &$T_3^2 S_3 T_3$   &$T_3 S_3 T_3^2$ &$T_3^2 S_3 T_3$ && $ X_4$ & $ X_4$ & $X_4$ \\
\bottomrule
\end{tabular}
\caption{Representations of $Q_8\rtimes A_4$ in the chosen basis. The first 4 representations are the unfaithful $A_4=\vev{S,T}$ representations to which the leptons are assigned (with $\rho(X)=\mathbbm{1}$). Note that the representations $\MoreRep{4}{i}$ are double valued, i.e. $\rho(Z(G)=X^2)=-\mathbbm{1}$, whereas the other representations are single valued ($\rho(X^2)=\mathbbm{1}$). $\MoreRep{1}{2,3}$ and $\MoreRep{4}{2,3}$ are complex, the other representations are real. \label{tab:Q8rtimesA4representations}}
\end{table}
\ra{1}

\subsection{Clebsch-Gordan Coefficients: Quartets}
The most important Clebsch-Gordan coefficients for the quartets $a,b\sim\MoreRep{4}{1}$ are given by:
\begin{align}
(a^\dagger b)_{\MoreRep{1}{1}}&=\frac{1}{2} \left(a^\dagger_1 b_1+a^\dagger_2 b_2+a^\dagger_3 b_3+a^\dagger_4 b_4\right)
\end{align}
and the triplets:
\begin{align}
(a^\dagger b)_{\MoreRep{3}{1}}&=\frac12\left(
\begin{array}{c}
-a^\dagger_4 b_1+a^\dagger_3 b_2-a^\dagger_2 b_3+a^\dagger_1 b_4\\
-a^\dagger_3 b_1-a^\dagger_4 b_2+a^\dagger_1 b_3+a^\dagger_2 b_4\\
a^\dagger_2 b_1-a^\dagger_1 b_2-a^\dagger_4 b_3+a^\dagger_3 b_4
\end{array}
\right) &
(a^\dagger b)_{\MoreRep{3}{2}}&=\frac12\left(
\begin{array}{c}
a^\dagger_4 b_1+a^\dagger_3 b_2+a^\dagger_2 b_3+a^\dagger_1 b_4 \\
a^\dagger_3 b_1+a^\dagger_4 b_2+a^\dagger_1 b_3+a^\dagger_2 b_4 \\
a^\dagger_2 b_1+a^\dagger_1 b_2+a^\dagger_4 b_3+a^\dagger_3 b_4
\end{array}
\right) \nonumber\\
(a^\dagger b)_{\MoreRep{3}{3}}&= \frac12\left(
\begin{array}{c}
a^\dagger_1 b_1-a^\dagger_2 b_2-a^\dagger_3 b_3+a^\dagger_4 b_4 \\
-a^\dagger_1 b_1+a^\dagger_2 b_2-a^\dagger_3 b_3+a^\dagger_4 b_4 \\
-a^\dagger_1 b_1-a^\dagger_2 b_2+a^\dagger_3 b_3+a^\dagger_4 b_4
\end{array}
\right) &
(a^\dagger b)_{\MoreRep{3}{4}}&= \frac12\left(
\begin{array}{c}
a^\dagger_4 b_1-a^\dagger_3 b_2-a^\dagger_2 b_3+a^\dagger_1 b_4 \\
-a^\dagger_3 b_1+a^\dagger_4 b_2-a^\dagger_1 b_3+a^\dagger_2 b_4 \\
-a^\dagger_2 b_1-a^\dagger_1 b_2+a^\dagger_4 b_3+a^\dagger_3 b_4
\end{array}
\right) \\\nonumber
(a^\dagger b)_{\MoreRep{3}{5}}&= \frac12\left(
\begin{array}{c}
-a^\dagger_4 b_1-a^\dagger_3 b_2+a^\dagger_2 b_3+a^\dagger_1 b_4 \\
a^\dagger_3 b_1-a^\dagger_4 b_2-a^\dagger_1 b_3+a^\dagger_2 b_4 \\
-a^\dagger_2 b_1+a^\dagger_1 b_2-a^\dagger_4 b_3+a^\dagger_3 b_4
\end{array}
\right)
\end{align}
Note that $\left[(a^\dagger b)_{\MoreRep{3}{i}}\right]^*=(b^\dagger a)_{\MoreRep{3}{i}} $ is real for $i=2,3,4$ and $\left[(a^\dagger b)_{\MoreRep{3}{i}}\right]^*=-(b^\dagger a)_{\MoreRep{3}{i}} $ for $i=1,5$. 

\subsection{Clebsch-Gordan Coefficients: Triplets}
Furthermore, the most important Clebsch-Gordan coefficients for the 3-dimensional representations \MoreRep{3}{i} are described by
\begin{align}
\label{eq:CGcoeffA4}
(a^\dagger b)_{\MoreRep{1}{1}}&= \frac{1}{\sqrt{3}}\left(  a_1^\dagger b_1 + a_2^\dagger b_2 + a_3^\dagger b_3  \right)\nonumber\\
(a^\dagger b)_{\MoreRep{1}{2}}&=\frac{1}{\sqrt{3}}\left(a_1^\dagger b_1+\omega^2 a_2^\dagger b_2+\omega a_3^\dagger b_3\right)&
(a^\dagger b)_{\MoreRep{1}{3}}&=\frac{1}{\sqrt{3}}\left(a_1^\dagger b_1+\omega a_2^\dagger b_2+\omega^2 a_3^\dagger b_3\right)\\\nonumber
(a^\dagger b)_{A,\Rep{3}}&=\frac 12 \left(\begin{array}{c} a_2^\dagger b_3-a_3^\dagger b_2\\a_3^\dagger b_1-a_1^\dagger b_3\\a_1^\dagger b_2-a_2^\dagger b_1\end{array}\right)&
(a^\dagger b)_{S,\Rep{3}}&=\frac 12\left(\begin{array}{c} a_2^\dagger b_3+a_3^\dagger b_2\\a_3^\dagger b_1+a_1^\dagger b_3\\a_1^\dagger b_2+a_2^\dagger b_1\end{array}\right)\;,
\end{align}
where $(a_1,a_2,a_3),\,(b_1,b_2,b_3)\sim \Rep{3}$. Note that $(a^\dagger a)_{A,\Rep{3}}$ is imaginary and $(a^\dagger a)_{S,\Rep{3}}$ is real.
Other important products are the product of $a\sim \MoreRep{3}{5}$ and $b\sim \MoreRep{3}{4}$:
\begin{align}
(a^\dagger b)_{\MoreRep{3}{1}}&=\left(\begin{array}{c} a_2^\dagger b_3\\a_3^\dagger b_1\\a_1^\dagger b_2\end{array}\right)&
(a^\dagger b)_{\MoreRep{3}{2}}&=\left(\begin{array}{c} a_3^\dagger b_2\\a_1^\dagger b_3\\a_2^\dagger b_1\end{array}\right)&
(a^\dagger b)_{\MoreRep{3}{3}}&=\left(\begin{array}{c} a_3^\dagger b_3\\a_1^\dagger b_1\\a_2^\dagger b_2\end{array}\right)\;,
\end{align}
of $a\sim \MoreRep{3}{5}$ and $b\sim \MoreRep{3}{2}$
\begin{align}
(a^\dagger b)_{\MoreRep{3}{1}}&=\left(\begin{array}{c} a_3^\dagger b_2\\a_1^\dagger b_3\\a_2^\dagger b_1\end{array}\right)&
(a^\dagger b)_{\MoreRep{3}{3}}&=\left(\begin{array}{c} a_2^\dagger b_2\\a_3^\dagger b_3\\a_1^\dagger b_1\end{array}\right)&
(a^\dagger b)_{\MoreRep{3}{4}}&=\left(\begin{array}{c} a_2^\dagger b_3\\a_3^\dagger b_1\\a_1^\dagger b_2\end{array}\right)
\end{align}
and of $a\sim \MoreRep{3}{4}$ and $b\sim \MoreRep{3}{2}$
\begin{align}
(a^\dagger b)_{\MoreRep{3}{1}}&=\left(\begin{array}{c} a_2^\dagger b_3\\a_3^\dagger b_1\\a_1^\dagger b_2\end{array}\right)&
(a^\dagger b)_{\MoreRep{3}{3}}&=\left(\begin{array}{c} a_1^\dagger b_1\\a_2^\dagger b_2\\a_3^\dagger b_3\end{array}\right)&
(a^\dagger b)_{\MoreRep{3}{5}}&=\left(\begin{array}{c} a_3^\dagger b_2\\a_1^\dagger b_3\\a_2^\dagger b_1\end{array}\right)\;.
\end{align}

%%%%%%%%%%%%%%%%%%%%%%%%%%%%%%%%%%%%%%%%%%%%%%%%%%%%%%%%%%%%%%%%%%%55
%%%%%%%%%%%%%%%%%%%%%%%%%%%%%%%%%%%%%%%%%%%%%%%%%%%%%%%%%%%%%%%%%%%55
%%%%%%%%%%%%%%%%%%%%%%%%%%%%%%%%%%%%%%%%%%%%%%%%%%%%%%%%%%%%%%%%%%%55

%\printbibliography
\bibliography{notes}
\end{document}